\documentclass[onecolumn]{aa} 

\usepackage{graphicx}
\usepackage{txfonts}
\usepackage{epsfig}			
\usepackage{graphicx,color}		
\usepackage{amssymb}			
\usepackage{color}			
\usepackage{url}			
\usepackage{amsmath}			
\usepackage{rotating}			
\usepackage{float}			
\usepackage{textcomp}			
\usepackage{psfig}
\usepackage{epstopdf}
\usepackage{dcolumn}
\usepackage{times}
\usepackage{tabularx}
\usepackage[english]{babel}
\usepackage[normalem]{ulem}
\usepackage{float}
\usepackage{ragged2e}
\usepackage{subfig}
\begin{document}

   \title{Kodaikanal Digitized White-light Data Archive (1921-2011): Analysis of various solar cycle features}

   \author{Sudip Mandal
          \inst{1}
          \and
          Manjunath Hegde
          \inst{1}
          \and
          Tanmoy Samanta
          \inst{1}
          \and
          Gopal Hazra
          \inst{1,2}
          \and
          Dipankar Banerjee
          \inst{1,3}
          \and
          Ravindra B.
          \inst{1}
          }

   \institute{Indian Institute of Astrophysics, Koramangala, Bangalore 560034, India.\\
              \email{sudip@iiap.res.in}  \\
         \and
             Department of Physics, Indian Institute of Science, Bangalore 560012, India  \\
         \and
       Center of Excellence in Space Sciences India, IISER Kolkata, Mohanpur 741246, West Bengal, India  \\
             }

   \date{Received ----; accepted -----}

 
  \abstract
   {Long-term sunspot observations are key to understand and predict the solar activities and its effects on the space weather.
   Consistent observations which are  crucial for long-term variations studies, 
   are generally not available due to upgradation/modifications of observatories over the course of time. We present the data for a period of
   90 years acquired from persistent observation  at the Kodaikanal observatory in India. 
   }
   {We aim to build a uniform sunspot area time series along with their positions, for the period of 90 years (1921-2011), as obtained from the newly digitized and calibrated white-light images from the Kodaikanal observatory. 
    Our aim is to compare this new time series with known sources and confirm some of earlier reported results with additional new aspects. 
    }
   {We use an advanced semi-automated algorithm to detect the sunspots form each calibrated white-light image. Area, longitude and latitude of each of the detected sunspots are derived. Implementation of a semi-automated method is very necessary in such studies as it minimizes the human bias in the detection procedure. }
   {Daily, monthly and yearly sunspot area variations obtained from the Kodaikanal, compared well with the Greenwich sunspot area data. 
   We find an exponentially decaying distribution for the individual sunspot area for each of the solar cycles. 
   Analyzing the histograms of the latitudinal distribution of the detected sunspots, we find Gaussian distributions, in both the hemispheres, with the centers at $\sim$15$^{\circ}$ latitude. 
   The height of the Gaussian distributions are different for the two hemispheres for a particular cycle. 
   Using our data, we show clear presence of Waldmeier effect which correlates the rise time with the cycle amplitude. Using the wavelet analysis, we explored different periodicities of different time scales present in the sunspot area times series.}
   {}


   \titlerunning{Kodaikanal Digitized White-light Data Archive (1921-2011)}
   \maketitle
  \justify
%

\section{Introduction}

 The Sun has a strong and complex magnetic field which is very dynamic and changes over various time scales.
 The magnetic activity govern the solar radiative energy, radio flux, CME’s, solar wind, heliospheric magnetic field and, consequently, cosmic ray flux at Earth.
 Sunspots are the enhanced magnetic patches on the solar photosphere and act as an excellent proxy for the eleven year solar cycle and the magnetic activities. 
 The number of sunspots observed on the solar surface varies periodically, going through strong successive maxima and minima with a periodicity around 11-year known as `sunspot cycle' or `solar cycle'.
 Sunspot observations in modern era have been started with Galileo Galilei using a telescope. 
 Though there are reports of sunspot observations from naked eyes earlier than that (see \citet{2007AdSpR..40..929V} for an exhaustive review on this). Sunspots which are clearly visible in white-light images, are recorded by different 
observatories around the world during different epochs of the last century. Royal Observatory of Greenwich (RGO) has a white-light data archive with observations stating from 1874 to 1976. This database is still being updated since 1976 from the Solar Optical Observing Network (SOON) observations. (RGO officially transferred the sunspot observing program to the Debrecen Heliophysical Observatory in Hungary near the end of 1976.)

 Kodaikanal Solar Observatory \citep{2010ASSP...19...12H} is observing the Sun since 1904. 
 This century long observation has been carried out in three different wavelength bands, 
 white-light (since 1904), Ca-K (since 1907) and in H-alpha (since 1912). 
 Among these three wavelengths, white-light observations were taken without any changes of the telescope optics since 1918. This makes the data set  very efficient for long term solar studies.  Kodaikanal white-light images have been used in the past by many authors to study the time evolution of different sunspot parameters. \citet{1993SoPh..146...27S} used the white-light data, for a period of 45 years (1940-1985), and compared their results with Mount Wilson white-light data. In the subsequent years, various aspects of the sunspot and their variation with the solar cycles have been studied using this data set. Solar rotation rates \citep{1999SoPh..186...25H,1999SoPh..188..225G}, axial tilt of the sunspot groups \citep{1999SoPh..189...69S}, tilt 
angle and size variation of the sunspot groups \citep{2000SoPh..196..333H}, rotational rate variation with the age of the sunspot groups \citep{2003SoPh..214...65S} are some of the highlighted works produced from this data set.

Sunspots have a strong link with large scale events like solar flares and CMEs which directly affects the space-weather and earth's atmosphere. Each sunspot cycle is unique in terms of the cycle amplitude and the duration. 
This makes it interesting to predict the forthcoming solar activities by analyzing a particular cycle. Long term solar studies are thus an unique opportunity 
to establish any empirical relations between different parameters of the solar cycles and enhance the accuracy of the prediction about the next cycles. 
In this paper we present the results from the digitized and calibrated Kodaikanal white-light sunspot data which covers the period 1921 to 2011. Section~\ref{sec_tel} and Section~\ref{sec_digi} 
briefly explain the details of the Kodaikanal white-light telescope and the digitization of the photographic plates whereas Section~\ref{sec_detect} describes the semi-automated algorithm 
used for sunspot detection. Different aspects of the sunspot area time series such as daily sunspot variation (Section~\ref{sec_area}), 
north-south asymmetry(Section~\ref{sec_ns}), latitudinal distributions (Section~\ref{sec_lat}) are presented in the subsequent sections. 
We discuss the correlations of the different cycle parameters with the cycle amplitude in Section~\ref{sec_wal} and explored different periodicities present in the sunspot area time series  in Section~\ref{sec_period}. The summary and conclusions are drawn in Section~\ref{sec_summary}.


\section{Telescope and Photographic plates}\label{sec_tel}

 Regular observations at Kodaikanal observatory started at 1904 using a white-light telescope with a 10-cm aperture lens and a f/15 light beam. 
  Between 1912 to 1917, the objective lens had been changed a few times and from 1918, the telescope has been installed with a 15-cm achromatic lens.
 This new configuration produced  a 20.4 cm size image of the Sun in the image plane. Photographic plates were used to capture the image. 
 The same telescope is being used since 1918 to 2011 (the observations with this telescope is continuing even today but those images have not been calibrated and thus they are not part of the data used in this paper) to take regular  white-light observation of the sun. More details on the telescope optics and its initial changes are described in \citet{1993SoPh..146...27S}.

  Initial white-light images, from 1904,  were stored in Ilford special Lantern photographic plates which had a size of 25.4 cm$\times$25.4 cm. In 1975, these plates became unavailable and thus the old plates had been replaced by high-contrast film of the same size. These photographic plates and the films, containing the solar images, have been stored carefully in individual paper envelopes at the obsevatory. Also a log book containing the observational details for each observation, was also preserved carefully. 

\section{Digitization and Calibration of the white-light images }\label{sec_digi}

  Normally the observations are taken in the morning before 10 IST when the seeing condition at the site is good (see \citet{1967SoPh....1..151B} for details of observing condition at Kodaikanal observatory). In most of the cases we have one image per day. For a better use of the scientific data, a new program for the high resolution digitization and calibration of the solar images recorded in the photographic plates and films, was initiated a few years back at Kodaikanal. The first representative results (for three solar cycles), from this digitization, was presented in \citet{2013A&A...550A..19R}. The raw images have been digitized by a 4k$\times$4k scientific grade CCD camera using a uniform source sphere digitizer unit. Calibration of the digitized data includes flat-fielding, conversion from emulsion density to relative density, identification of solar radius and the Sun center along with rotation correction (see \citet{2013A&A...550A..19R} for details of each process). Calibrated white-light data covers the period from 
1920 
to 2011. This includes solar cycle 16 to cycle 24 (still ongoing). On an average, 290 plates per year have been acquired. Here we must emphasize the fact that though the optics was unchanged for the past 90 years, but the image quality may have gone thorough several changes over this period due various reasons such as variable seeing, change of storage device from photographic plates to films and so on. These aspects of the time series for the Kodaikanal white-light digitized images have been studied thoroughly by \citet{2013A&A...550A..19R}.

We would also like to mention that for the period of 1905 till 1920, there are some issues with the raw images such as two perpendicular lines on the top of the images (one for N-S and another for E-W line of the Earth) and different image sizes. These pose difficulties in correctly orienting the images. This part of the calibration of 15 years, starting from 1905 to 1920, could not be handled by the current version of the code and has to be handled separately, which we hope to take up in near future.


  \begin{table*}[!htbp]
  \caption{\label{specifications} Specifications of Kodaikanal white-light images}
  	\centering
  	\begin{tabular}{lccc}
  		\hline
  		$Parameter$ & $Description$ \\ 
  		\hline
  		Observation period & 1904 to 2011 \\
  		Telescope & 15 cm objective with f/15 beam\\
  		Original storage & Plates/Films \\
  		Total number of plates & 31800 (1904 to 2011)\\
  		Average number of plates per year & 290\\
  		Calibration period & 1921 to 2011 \\
  		Final image size & 4k$\times$4k \\
  		Pixel scale & 0.62$''$ \\
  		Available image format & FITS and JPG \\
  		Raw images available at & \url{http://kso.iiap. res.in}\\
  		\hline
  	\end{tabular}
  	\label{table}
  \end{table*}

\section{Sunspot Detection}\label{sec_detect}

\begin{figure*}[!htbp]
\centering
\includegraphics[angle=90,width=1.0\textwidth]{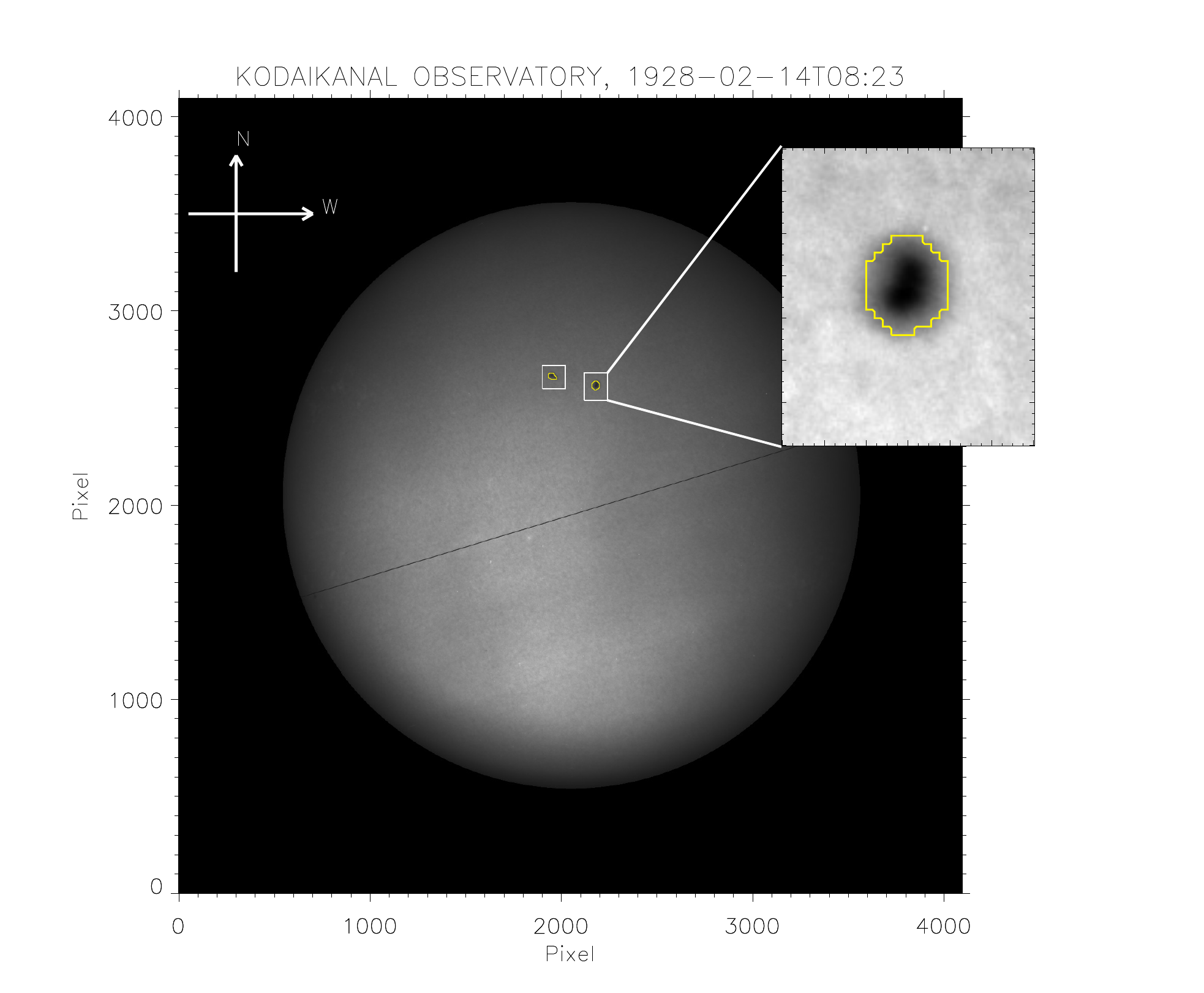}
\caption{ A representative image from the Kodaikanal white-light digitized data. Two detected sunspots are highlighted by two white rectangular boxes. In the inset we show a zoomed-in view of one of the sunspots. The yellow contours represent the detected area for the sunspots.}
\label{rep_image} 
\end{figure*}
 Sunspot area varies over the different epoch of a solar cycle. This increment and decrement of the sunspot area with the solar maxima and minima has been analyzed in the past. In this paper, for the first time, we are presenting the sunspot area variation for nearly about 90 years using Kodaikanal white-light digitized data. To get the variation in the sunspot area with the solar cycle, we have detected the sunspots using a semi-automated algorithm. 

  We have used a modified version of the `sunspot tracking and recognition algorithm' (STARA, \citet{2009SoPh..260....5W}) to identify the sunspots on our calibrated white-light images. This modified version of the code uses two intensity thresholds, in contrast to a single thresholding used in the original code. The working principle of this code is explained in details in \citet{2013A&A...550A..19R}. Although this procedure detects the sunspots automatically but it occasionally picks up dirt and the east-west line in the image. These artifacts have been removed manually by checking every contour drawn on the detected spots and using a cursor to rejects those which does not belong to a sunspot. Also, we have modified the code to get the position information (longitude and latitude from the `center of gravity' method) of the detected sunspots which was not implemented in the code used in \citet{2013A&A...550A..19R}.

  Figure~\ref{rep_image} shows a representative image of Kodaikanal white-light data where the detected sunspots are shown with the contours overplotted on them. In the inset of Figure~\ref{rep_image}, we show a zoomed-in view of one of the sunspots in the image. In this zoomed-in image, we notice the umbral and the penumbral area boundaries along with the contour which covers the whole-spot area perfectly. We will refer the `whole-spot' area as the `sunspot area' hereafter.
\section{Results}

\subsection{Sunspot area variation}\label{sec_area}

\begin{figure*}[!htbp]
\centering
\includegraphics[clip,width=1.\textwidth]{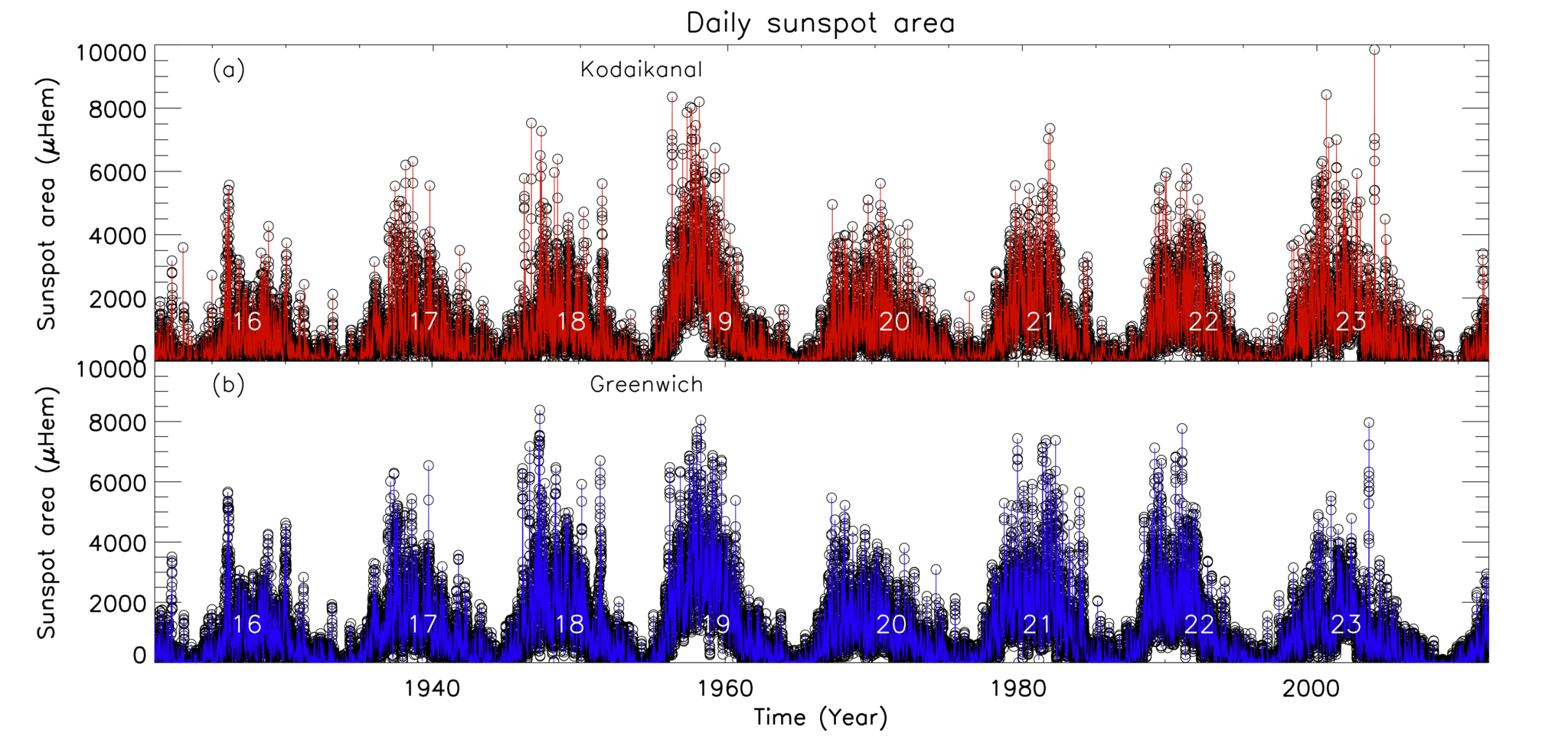}
\caption{ Panel (a) shows the daily sunspot area from the Kodaikanal white light images. In panel (b) we plot the same but for the Greenwich data. Individual sunspot cycle numbers are printed on both the panels.}
\label{daily_image} 
\end{figure*}

\begin{figure*}[!htbp]
\centering
\includegraphics[angle=90,clip,width=.95\textwidth]{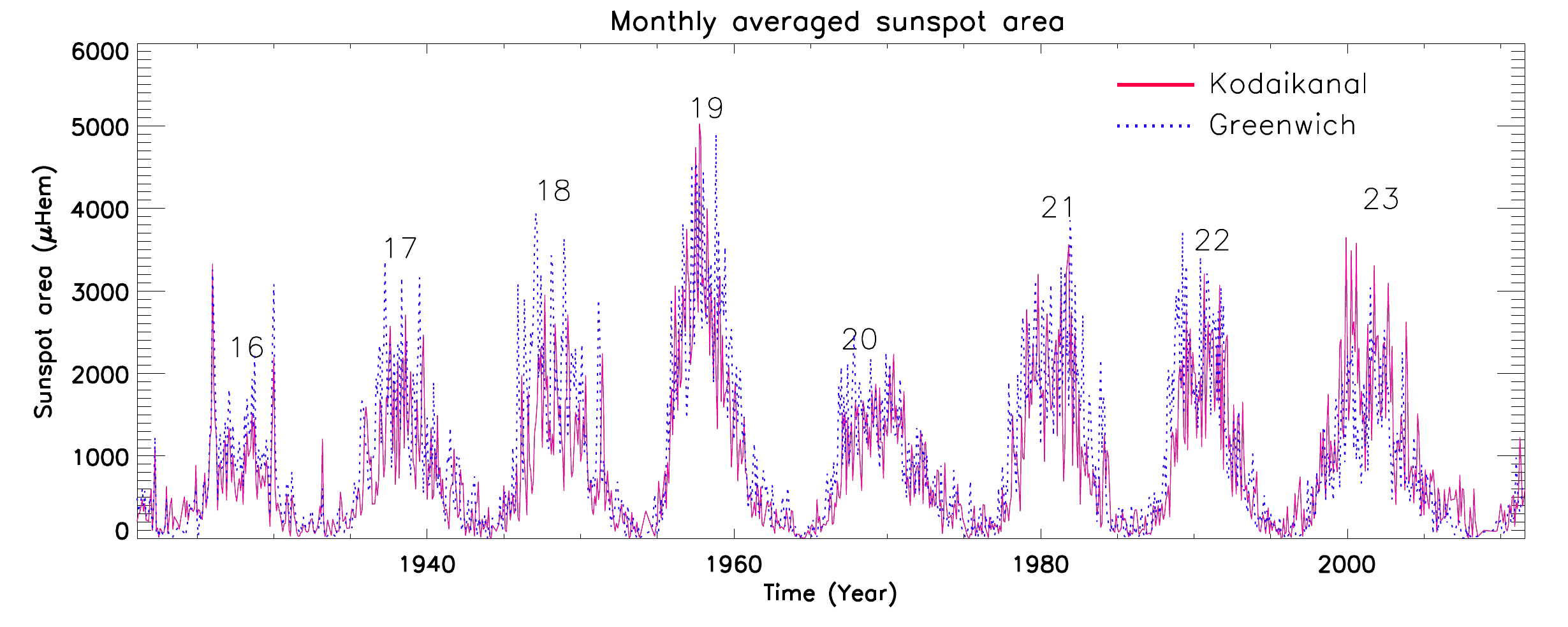}
\caption{ Monthly averaged sunspot data obtained from Kodaikanal (plotted in red) along with the data from Greenwich for the same period (plotted in blue).  }
\label{monthly_image} 
\end{figure*}

\begin{figure*}[!htbp]
\centering
\includegraphics[angle=90,clip,width=.95\textwidth]{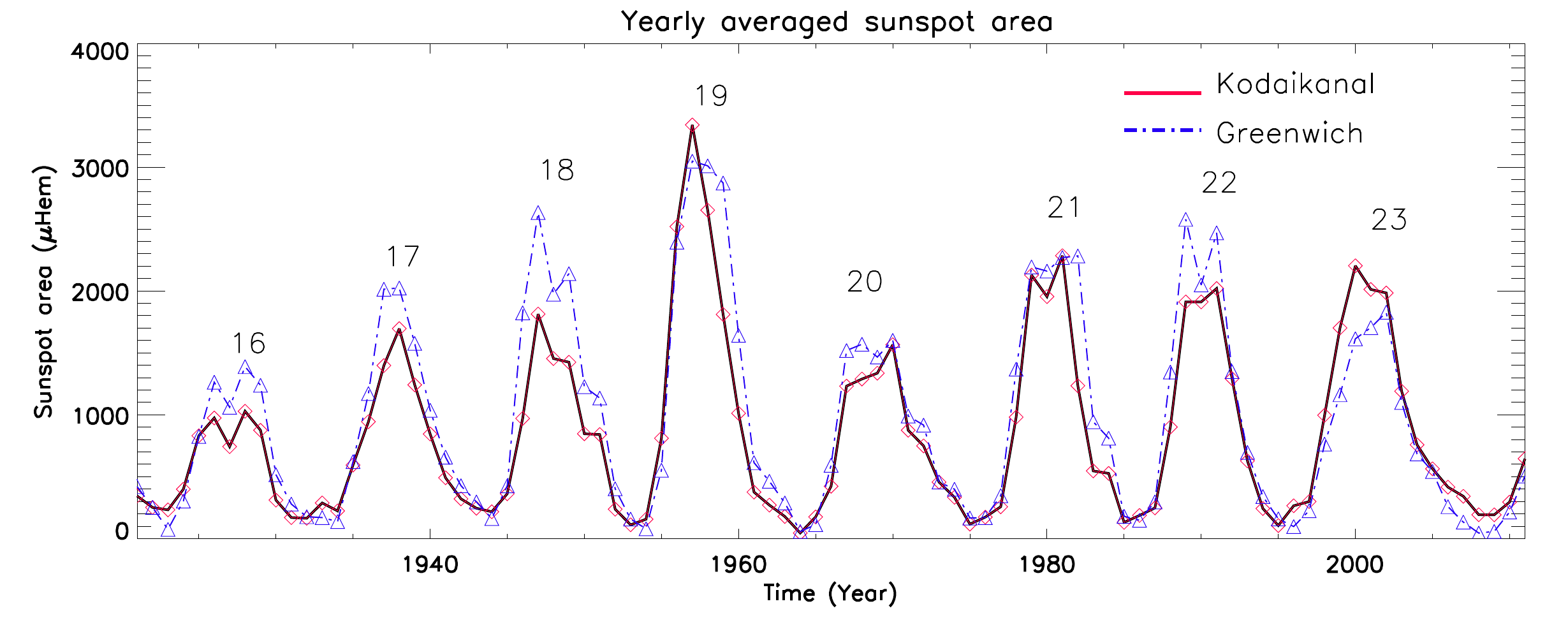}
\caption{ Yearly averaged sunspot area from Kodaikanal and Greenwich plotted in red and blue respectively. }
\label{yearly_image} 
\end{figure*}

   Using the above described method we have detected the sunspots using the Kodaikanal white-light images within a period from 1921 to 2011. For every detected sunspot, we have the area, longitude and latitude information as an output from our modified detection algorithm. Panel (a) in Figure\ref{daily_image} shows the daily sunspot area obtained from Kodaikanal. From the plot we immediately recover the 11-year sunspot cycle. Also, we notice that the individual cycle amplitudes vary over the entire time series.  Now to compare our results, we have used the Greenwich sunspot area data (plotted in panel (b) in Figure~\ref{daily_image}) which is available online at \url{http://solarscience.msfc.nasa.gov/ greenwch.shtm}. We see a very good match of our data with the Greenwich time series and hence validate our detection method as well as establishes the quality of the data.
 The time variation of the monthly averaged sunspot area obtained from the Kodaikanal is plotted in red in Figure~\ref{monthly_image}. The Greenwich data, for the same, is overplotted in blue. The fluctuations seen in the monthly averaged data, get further smoothed out in the yearly averaged plot shown in red in Figure~\ref{yearly_image}. From the overplotted Greenwich data (in blue) we notice a good overall match between the two data sets. There are certain differences also such as some of the double peak behaviors seen in Kodaikanal data are not so prominent in Greenwich and vice-versa. Also we see that the maximum difference of the amplitudes among the two data occur at the time of cycle maxima and the Greenwich data has higher area values (at the solar maximum) compared to the Kodaikanal data, in most of the cycles. For example from Figure~\ref{yearly_image} we notice that, near the time of maxima of cycle 18, Kodaikanal area values are significantly lower than the Greenwich values, though the pattern remains the same for the two data set. Because of poor photographic plate conditions, the semi-automated code did not detect all the spots around this cycle maxima, which results in an underestimation. We should also mention here that the Greenwich data has been cataloged using sunspot observations from different observatories around the globe (Cape of Good Hope, Kodaikanal and Mauritius) to incorporate for the missing days or bad data. Thus it includes different scaling in-order to incorporate these different data from various observatories. This may also lead to slight differences in values in comparison to the Kodaikanal data which has been complied from a single observatory for the whole 90 years of data duration. To improve the statistics further, we will compare original images from different sources in our forthcoming studies.
 This involves one to one cross calibration of the sunspot images from the Kodaikanal and the Greenwich observatories (on the simultaneous observational days). Such a study will also help towards generating a consistent sunspot area time series which can be used by the solar community.

A further comparison and relevant discussion of the Kodaikanal data with the Debrecen data base has been added in Appendix:B .

\begin{figure*}[!htbp]
\centering
\includegraphics[angle=90,width=0.8\textwidth]{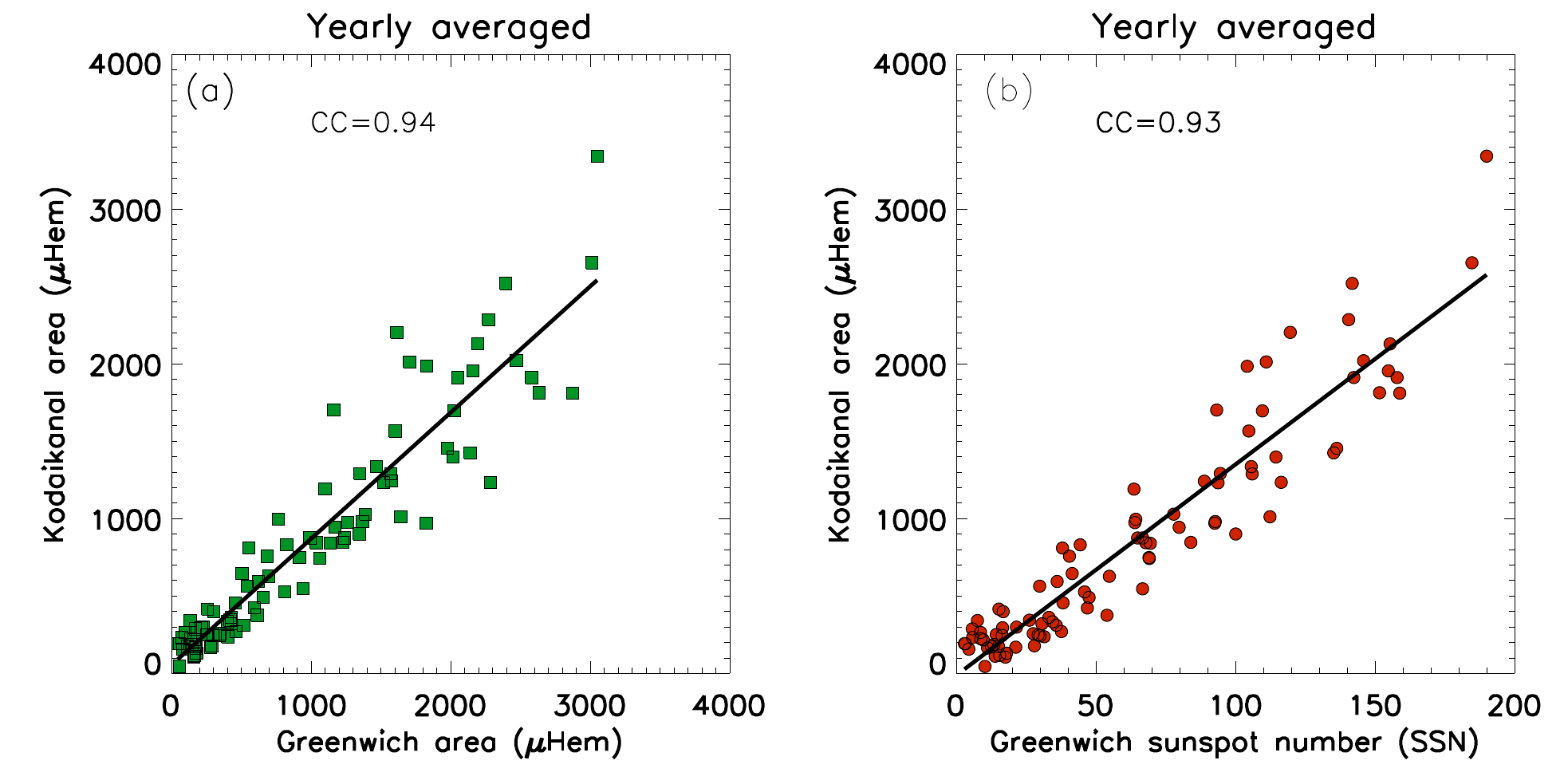}
\caption{ Panel (a) shows the scatter plot between the yearly averaged Kodaikanal and Greenwich area measurements. Panel (b) shows the scatter plot using the yearly averaged smoothed sunspot number (SSN) from Greenwich along with the obtained Kodaikanal sunspot area. Correlation coefficients (CC) are printed on both the panels.}
\label{compare} 
\end{figure*}

To quantify the correlation between the Kodaikanal and the Greenwich data, we plot the scatter diagram, in panel (a) in Figure~\ref{compare}, for the yearly averaged area values. The correlation coefficient, obtained in this case, is 0.94. We also plot a similar 
scatter diagram for the sunspot number, obtained from Greenwich data along with Kodaikanal sunspot area and we obtain a correlation coefficient of 0.93. Here we must emphasize that such a high correlation of the sunspot area with the sunspot number shows the goodness of use of sunspot area as an alternate proxy for the solar cycle.

\begin{figure*}[!htbp]
\centering
\includegraphics[width=0.8\textwidth]{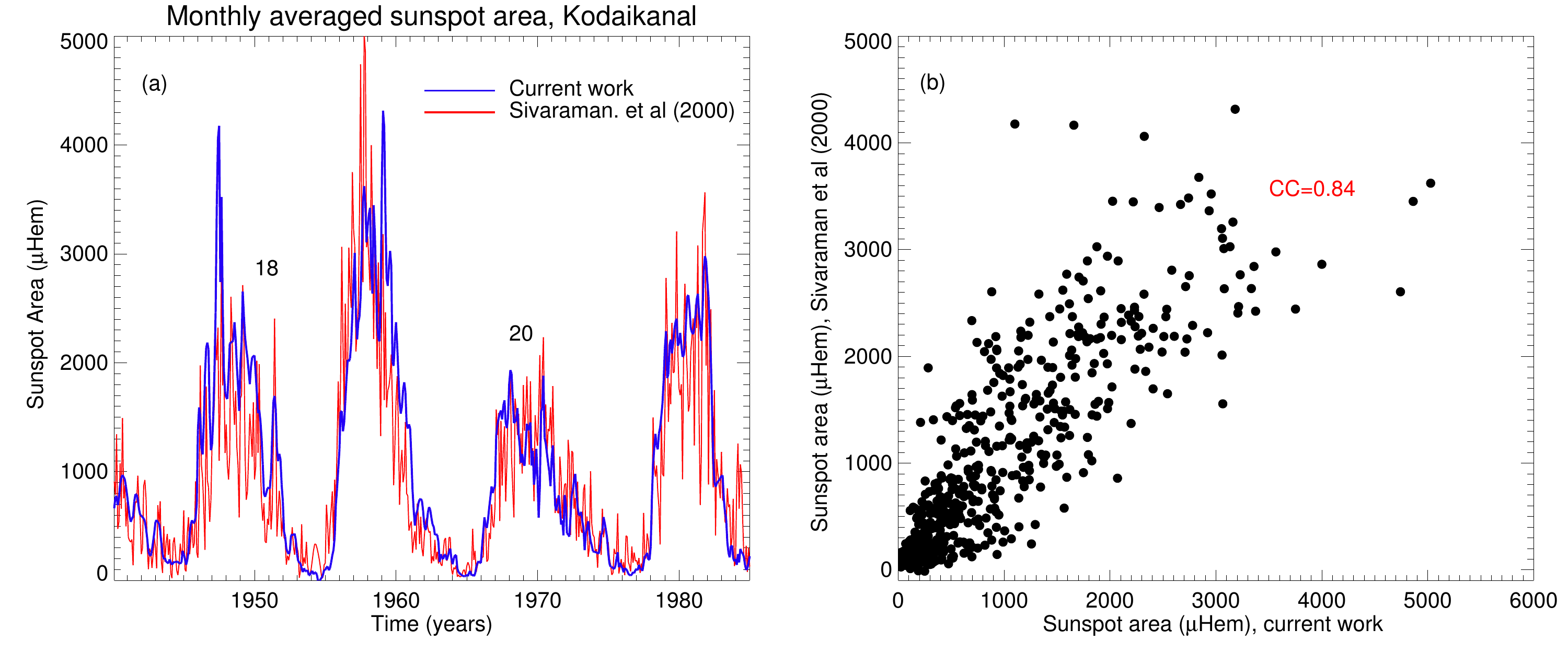}
\caption{ Panel (a) shows comparison between the manual (red line) and the automatic (blue line) detection of sunspot area from Kodaikanal white-light images for a period of 1940-1985. The scatter plot of these two area values, is shown in panel (b.)}
\label{manual_auto} 
\end{figure*}

 To validate our detection algorithm further, we compare our results with \citet{2000JApA...21..149S}. 
 These authors calculated the sunspot area from the Kodaikanal white-light images by using hand drawn contours. Here we must mention the fact that these authors did not provide the data in the tabular format in their paper. 
 We used the IDL routine \textit{read\_jpeg.pro} to extract the information from the Figure 2 in \citet{2000JApA...21..149S}. 
 In panel (a) in Figure~\ref{manual_auto} we plot monthly averaged sunspot area, during the period 1940-1985, for both the cases. A careful observation of the plot reveals that there are three notable deviations at three different months. We have critically checked our analysis at those locations and did not find over-estimation or under-estimation of the area values. As mentioned earlier, there is no written records of the previous work and thus it became impossible for us to comment on the results obtained by \citet{2000JApA...21..149S}. Also, in panel (b) we show the scatter plot between the two area values and we obtain a correlation of 0.84. This relatively high correlation value indicates the goodness of our semi-automated detection algorithm.

\subsection{Sunspot area distribution}

\begin{figure*}[!htbp]
\centering
\includegraphics[angle=90,width=1.\textwidth]{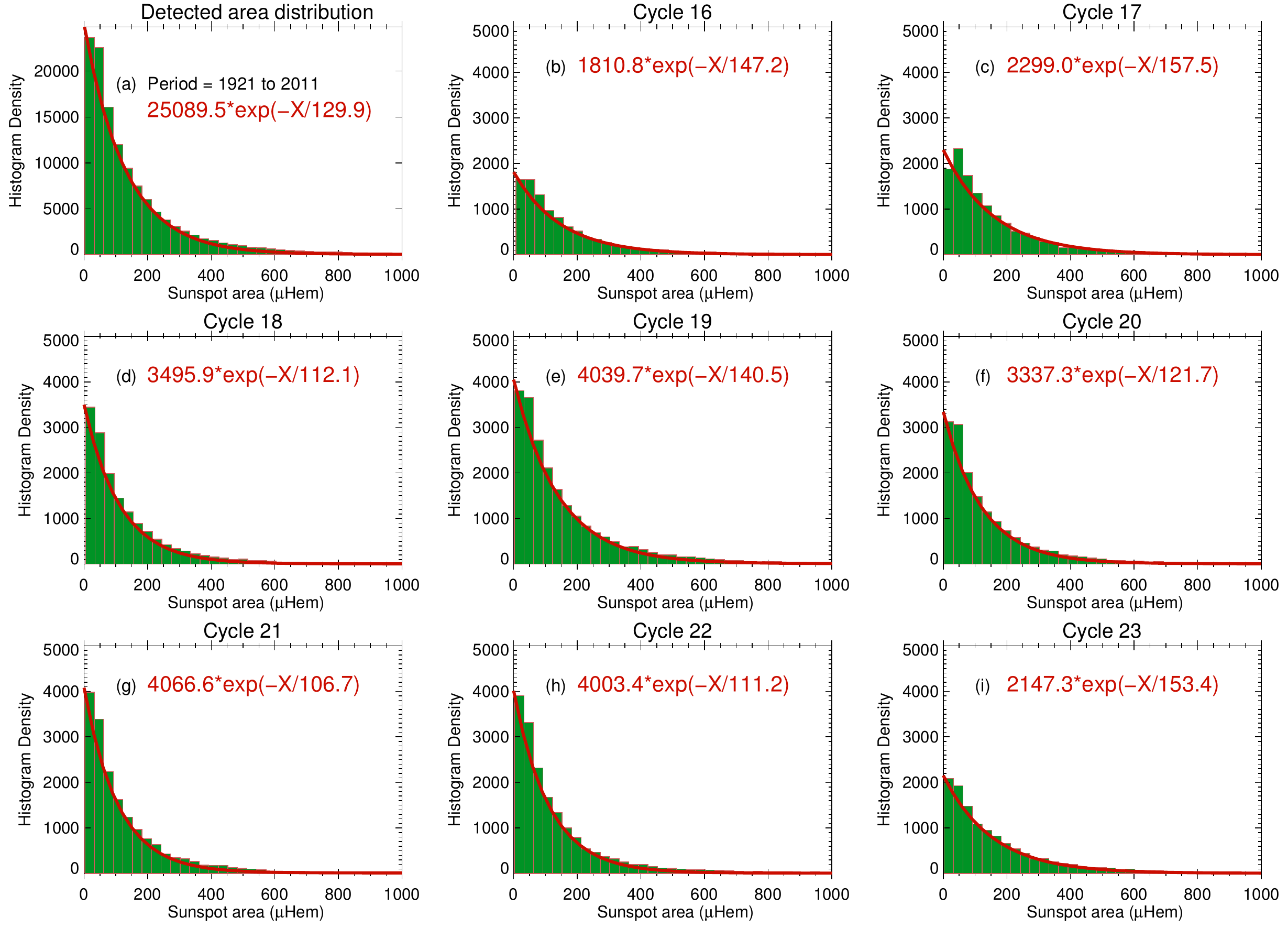}
\caption{ Plot showing the histograms (with a bin-size of 30 $\mu$Hem) of the individual area values (sizes) obtained from Kodaikanal white-light images. The histogram for the full data set is shown in panel (a). Panels (b-i) show the area distributions for the individual solar cycles ( cycle 16 to 23). }
\label{area_hist} 
\end{figure*}

Sunspots come in variety of sizes and it also vary with the different phases of a solar cycle. 
We investigate the size distribution using each sunspot area obtained from the Kodaikanal white-light images, for the entire period as well as for individual solar cycles. 
In panel (a) of Figure~\ref{area_hist}, we plot the histogram of the area values of each sunspots for the full data. We see a decaying exponential pattern in the histogram profile for 
the whole data set (panel (a)) as well as for individual cycles (panel (b-i)). We fit the histograms with a decaying exponential function of the form, Y=A$_0\exp^{-(\frac{X}{B})}$,
as shown by a red solid line in each panel. Analyzing the histograms for individual cycles, we notice that the number of large sunspots has a strong correlation with the overall cycle amplitude i.e the stronger cycles (e.g cycle 19) have more number of large sunspots compared to the weaker cycles (e.g cycle 16). The smaller sunspots (area $<$ 30 $\mu$Hem) do not seem to follow any certain 
rule. Cycle 19 and cycle 21 has different cycle amplitudes but the number of small sunspots are almost the same whereas the cycle 22 and cycle 23 has similar amplitudes but the number of small sunspots in this case is almost half than the former one.

\begin{figure*}[!htbp]
\centering
\includegraphics[angle=90,width=.8\textwidth]{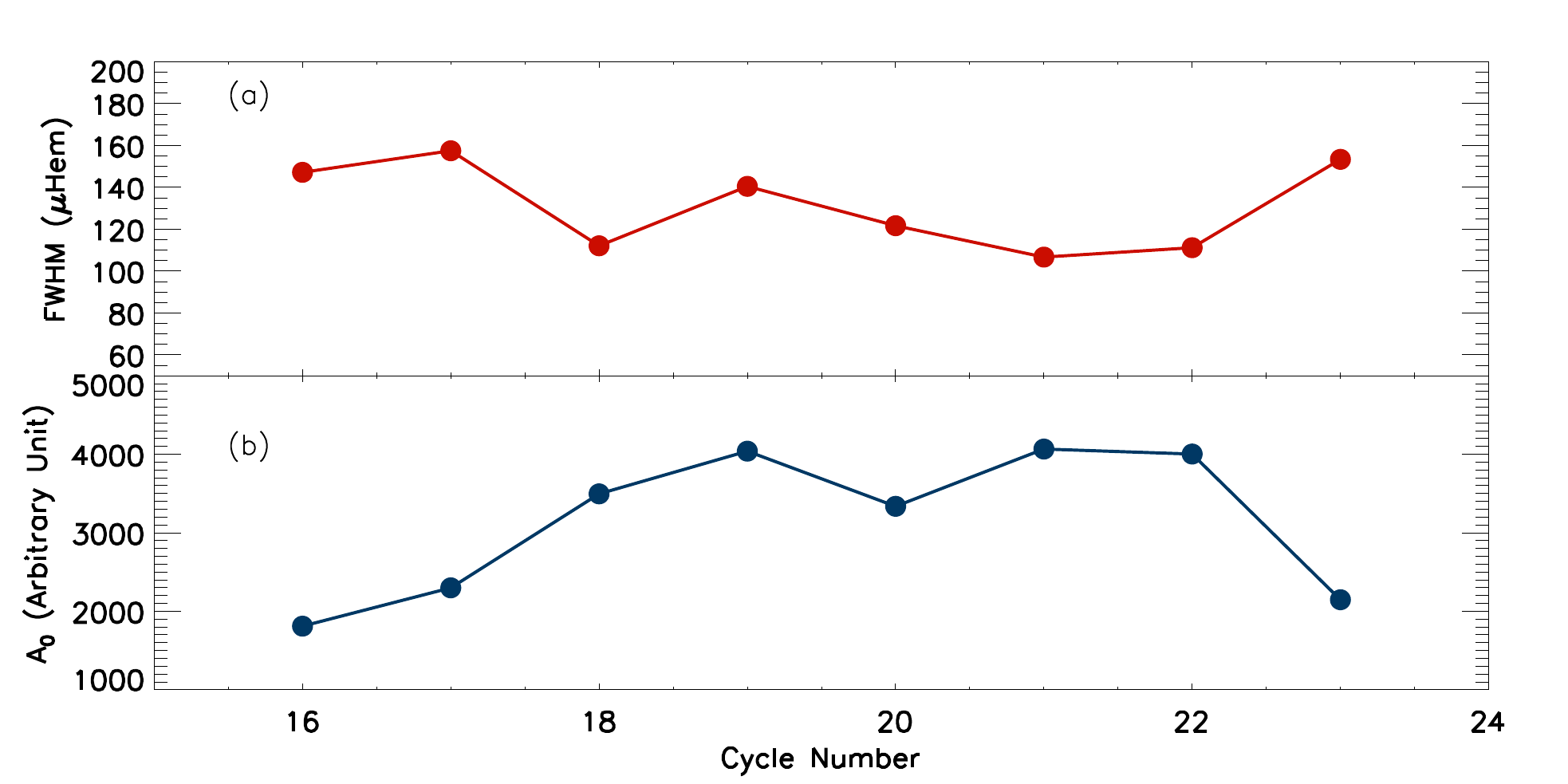}
\caption{ Figure shows the variations of the parameters,  FWHM (in panel a) and A$_0$ (in panel b), from the fitted exponential functions. }
\label{area_deviation} 
\end{figure*}

 In panel (a) and panel (b) of Figure~\ref{area_deviation}, we plot the full width at half maximum (FWHM) and the A$_0$ values obtained from the fitted curves from each of the histograms. The FWHM here is defined as the value where the amplitude drops to $1/\mathrm{e}$ times of A$_0$. We do notice some periodicity in the A$_0$ values but our data series is not sufficiently long to conclude any significant number. Also we notice a weak anti-correlation between the two parameters, A$_0$ and FWHM. 

\begin{figure*}[!htbp]
\centering
\includegraphics[width=1\textwidth]{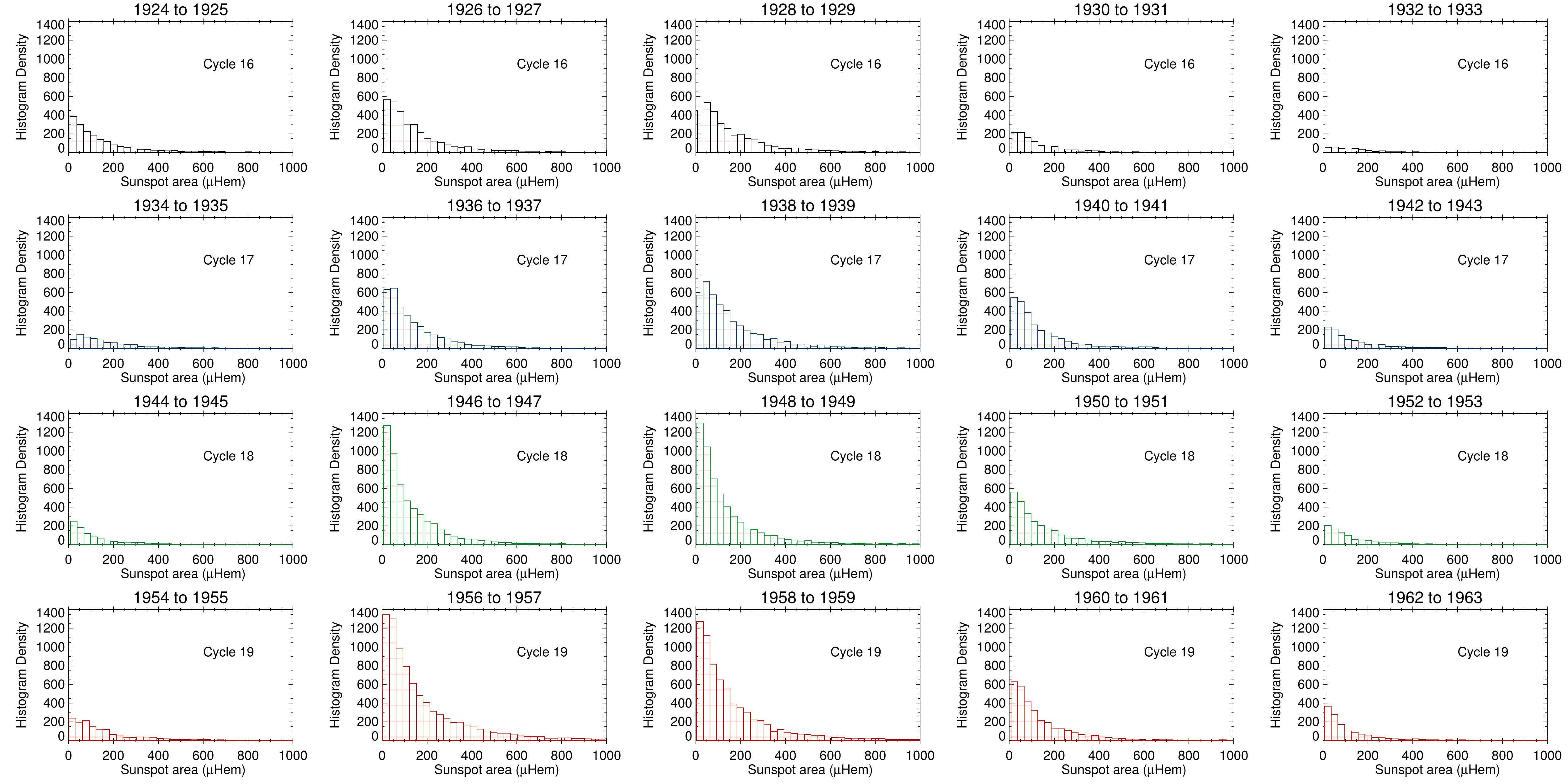}

\includegraphics[width=1\textwidth]{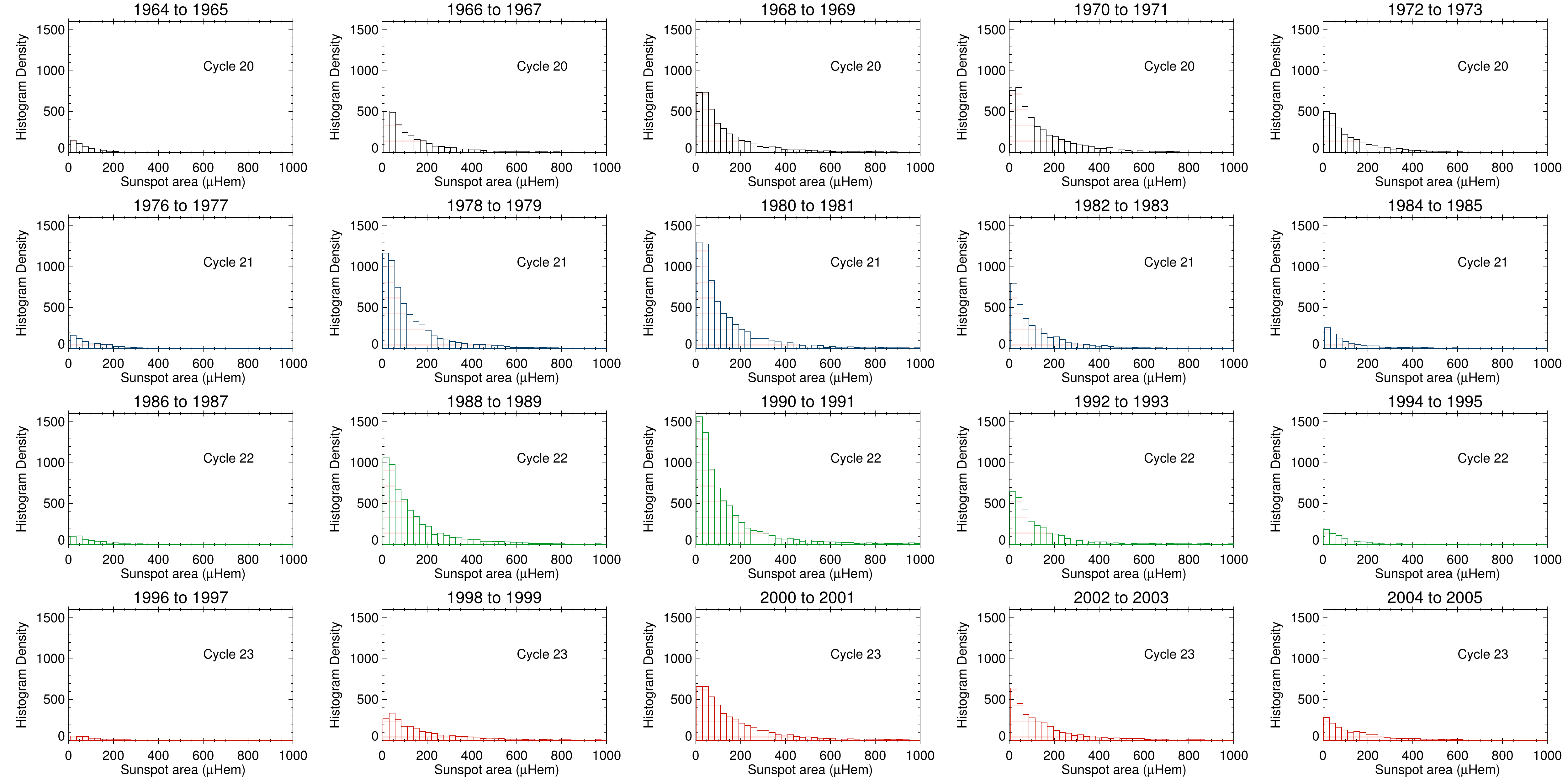}
\caption{ Figure shows the sunspot size distribution at different phases of the solar cycle for each individual cycles (cycle 19-23). The bin-size is 30 $\mu$Hem}
\label{phase2} 
\end{figure*}

In Figure~\ref{area_hist} we have shown that the individual sunspot area distribution for each cycle, is following an exponentially decreasing function. Next we investigate the same but for each phase of an individual cycle. We divided each cycle into two years bin and plotted the area distribution for each cycles in Figure~\ref{phase2}. We notice that the exponential trend remains in all the phases of a cycle. Though, at the time of cycle maxima, the number of smaller as well as bigger sunspots increases but the distribution remains the same. Another important aspect is that not only the number of big sunspots increases during the maxima of the cycle, but the number of the sunspots with smaller sizes also increases.


 We also fit these area histograms with a lognormal function and found a good match with the earlier results. These plots and their descriptions are given in Appendix~A.
\subsubsection{North-South area distribution}\label{sec_ns}

\begin{figure*}[!htbp]
\centering
\includegraphics[angle=90,width=0.8\textwidth]{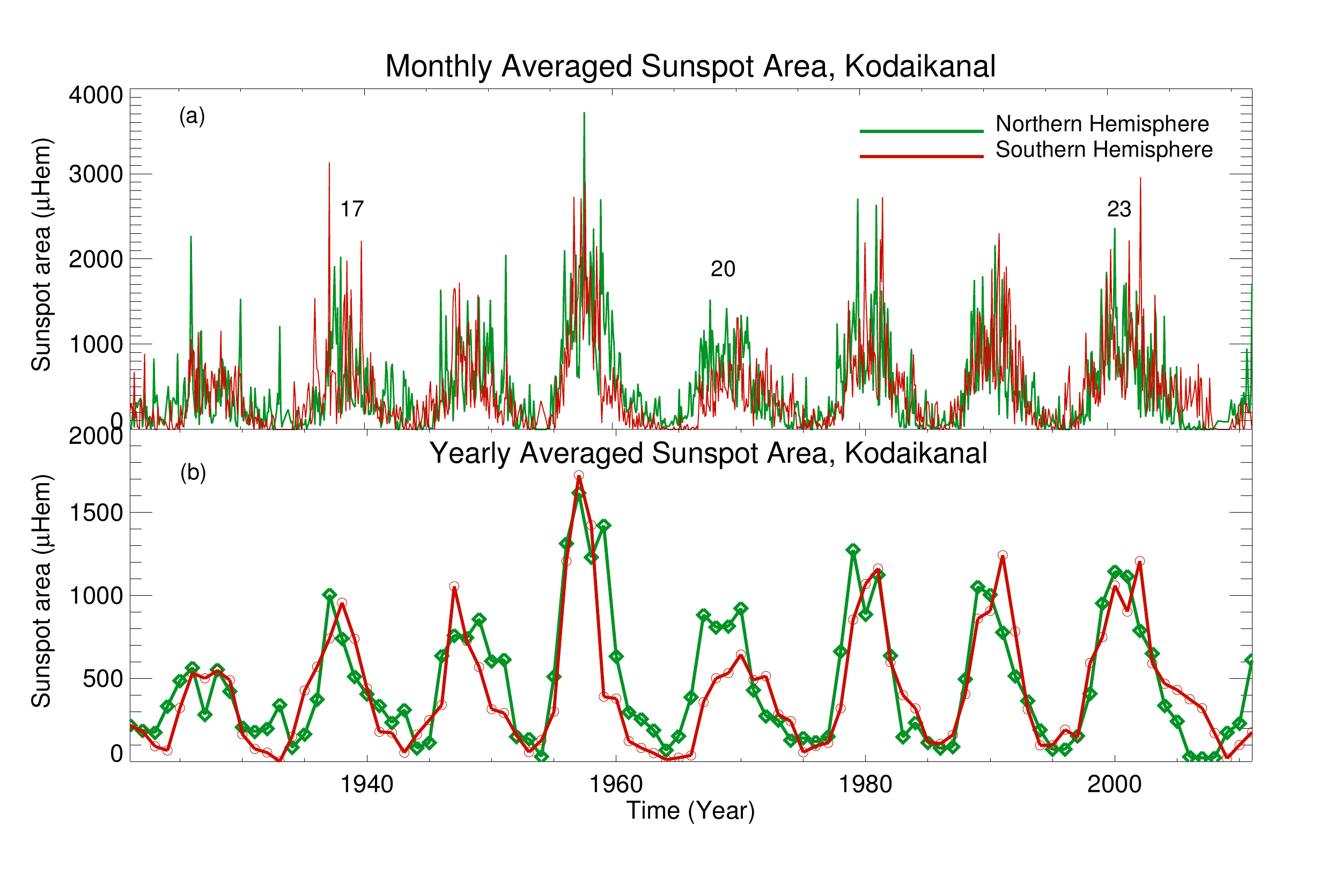}
\caption{ Panel (a) shows monthly averaged sunspot area variation with time for the northern (shown in green) and southern (shown in red) hemisphere. Panel (b) shows the same as previous but for yearly averaged data.}
\label{ns1} 
\end{figure*}

   Sunspots do not appear symmetrically in both the solar hemispheres and this phenomena is popularly known as `North-South' (N-S) asymmetry. This asymmetric behaviour in the sunspot area has been studied in the past \citep{1993A&A...274..497C,1994SoPh..152..481O,2005A&A...431L...5B,2011NewA...16..357B}.
We re-investigate the N-S asymmetry using our sunspot data extracted from the Kodaikanal white-light images. In panel (a) of Figure~\ref{ns1} we plot the time variation of the monthly averaged sunspot area for the individual hemispheres ( in green and red for northern and southern hemisphere respectively). 

We also plot the yearly averaged sunspot area for each hemispheres in panel (b). A closer inspection on panel (b) reveals that the double peak behavior is very often restricted to one hemisphere only. In the cycle 19 we see a clear double peak in the northern hemisphere whereas no such behavior is seen in the southern hemisphere for that cycle. Same behavior can bee seen for the cycle 21. In cycle 22 and 23, it is the southern hemisphere which shows the double peaks.

From the plots we see the sunspot area evolution with the solar cycle is not symmetric in its two hemispheres. There are different ways to quantify this asymmetricity. 
The two most popular methods are `Absolute asymmetry' parameter, defined as $A_{a}=\mathrm{(N-S)}$
and the `Normalized asymmetry' parameter defined as
$N_{a}=\mathrm{(N-S)/(N+S)}$
where $\mathrm{N}$ and $\mathrm{S}$ are the sunspot area in the northern and southern hemispheres.

\begin{figure*}[!htbp]
\centering
\includegraphics[angle=90,width=0.8\textwidth]{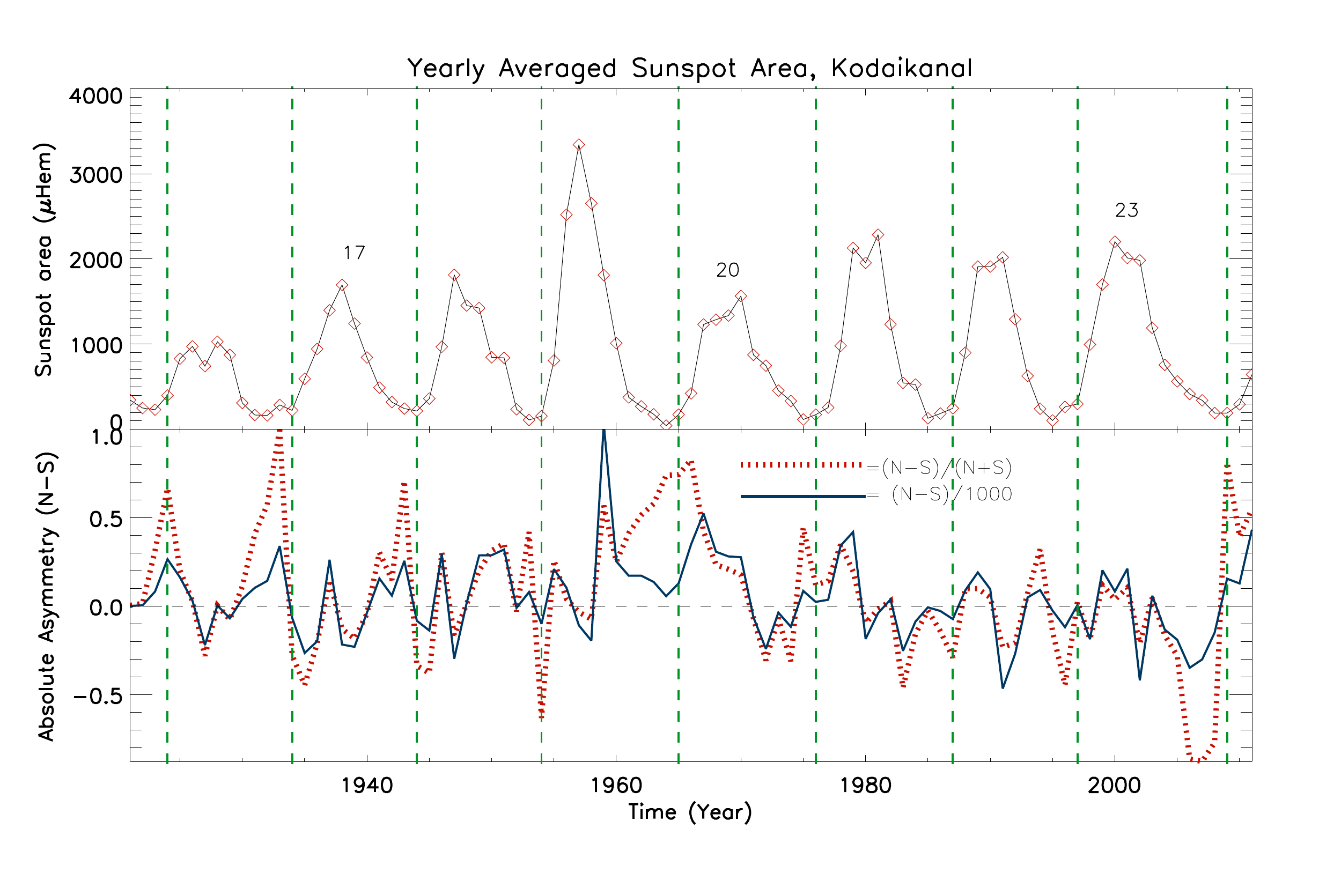}
\caption{ Top panel shows the yearly averaged sunspot area from Kodaikanal. In the bottom panel, we compare the two parameters used to compare the north-south asymmetry on the yearly averaged data. The normalized asymmetry (N-S)/(N+S) is shown by red dotted line whereas the blue solid line represent the absolute asymmetry parameter.}
\label{ns2} 
\end{figure*}
We compare the values of A$_a$ and N$_a$ for our dataset. In top panel of Figure~\ref{ns2} we plot the yearly averaged sunspot area where as in the bottom panel of ~\ref{ns2} we plot the A$_a$ in blue solid line and N$_a$ in red dotted line. We notice that the two parameters evolve differently with the progress of the solar cycle. The N$_a$ reaches its maximum value at the time of solar minima whereas the A$_a$ peaks at the solar maxima of each cycle. Such variation of the parameters can be explained by the fact that at the time of solar minimum there are very few spots appear at the surface of the sun, making the value of the N$_a$ very high, close to +1 or -1, depending on the appearance of the sunspot in the hemispheres. The advantage of use of A$_a$ over the N$_a$, specially for time series analysis, has been studied thoroughly in past \citep{2005A&A...431L...5B,2006A&A...447..735T}.
\subsubsection{Butterfly diagram and latitudinal area distribution}\label{sec_lat}

\begin{figure*}[!htbp]
\centering
\includegraphics[width=.95\textwidth]{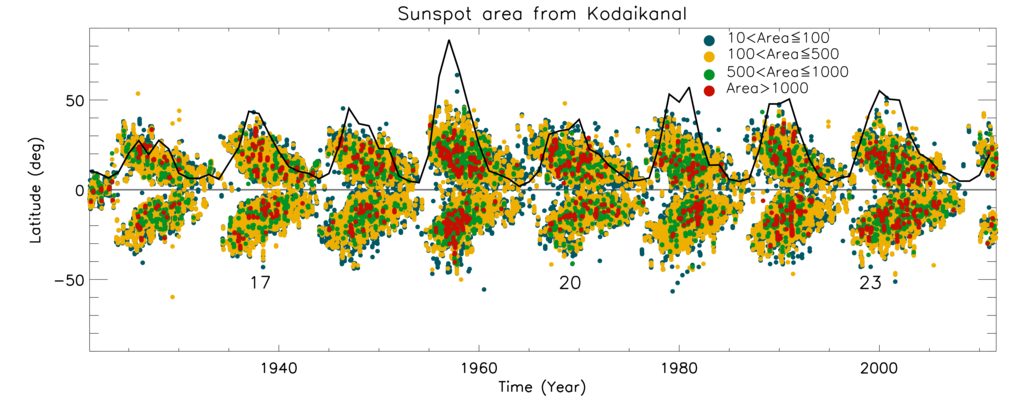}
\caption{ The butterfly diagram created using the individual sunspot area values (sizes) and their locations obtained from Kodaikanal white-light images. Cycle numbers are also printed in the plot. }
\label{butter} 
\end{figure*}

   In this section we explore the latitudinal variation of the emergent sunspots over the solar cycle evolution. In Figure~\ref{butter} we plot the time evolution of the latitude values of the detected sunspots from Kodaikanal. We also divided the sunspots into four size bins and overplotted them with different color on the same plot (different size bins are printed on the plot). 

 Figure \ref{butter} represents the  `butterfly diagram' which shows that the sunspots appear at higher latitude in the beginning of the solar cycle and then the activity belt drifts towards the equator as the cycle progress. We also overplot the yearly averaged sunspot area (scaled) by a solid black line. A careful observation on figure also reveals that the larger spots (having area~$\geq$ 1000 $\mu$Hem) appear at the time of solar maximum. Also we see that the number of these bigger sunspots depends on the strength of the cycle i.e the strongest cycle, cycle 19, has more number of spots bigger than 1000 $\mu$Hem compared to a weak cycle, cycle 16.

\begin{figure*}[!htbp]
\centering
\includegraphics[angle=90,width=0.95\textwidth]{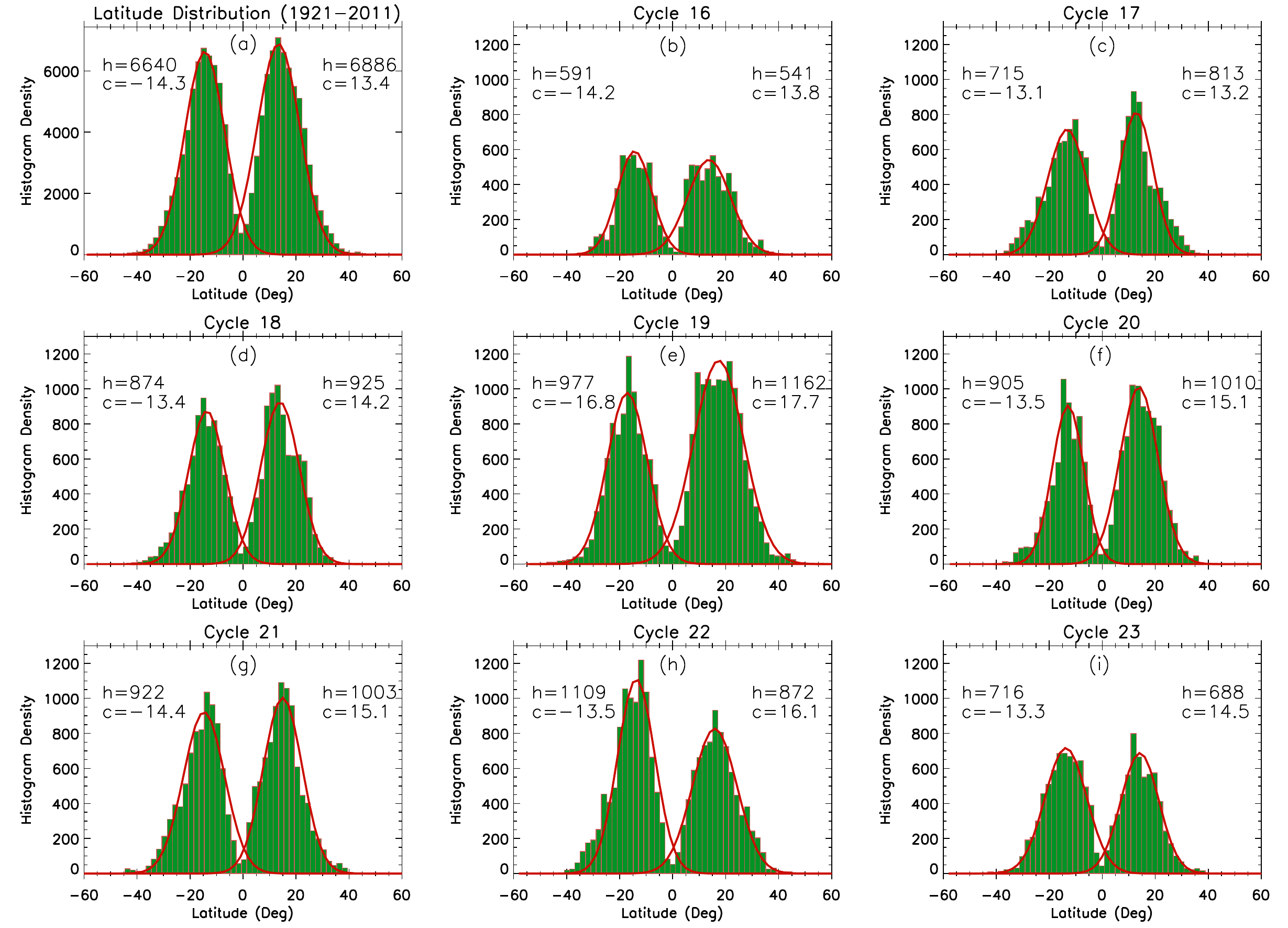}
\caption{ Histograms of the latitude distribution of the individual sunspot area values (sizes), from the Kodaikanal, are shown for both data set and for individual cycles. Fitted Gaussian in each histogram, is overplotted in red. }
\label{lat_distri} 
\end{figure*}

\begin{figure*}[!htbp]
\centering
\includegraphics[angle=90,width=0.95\textwidth]{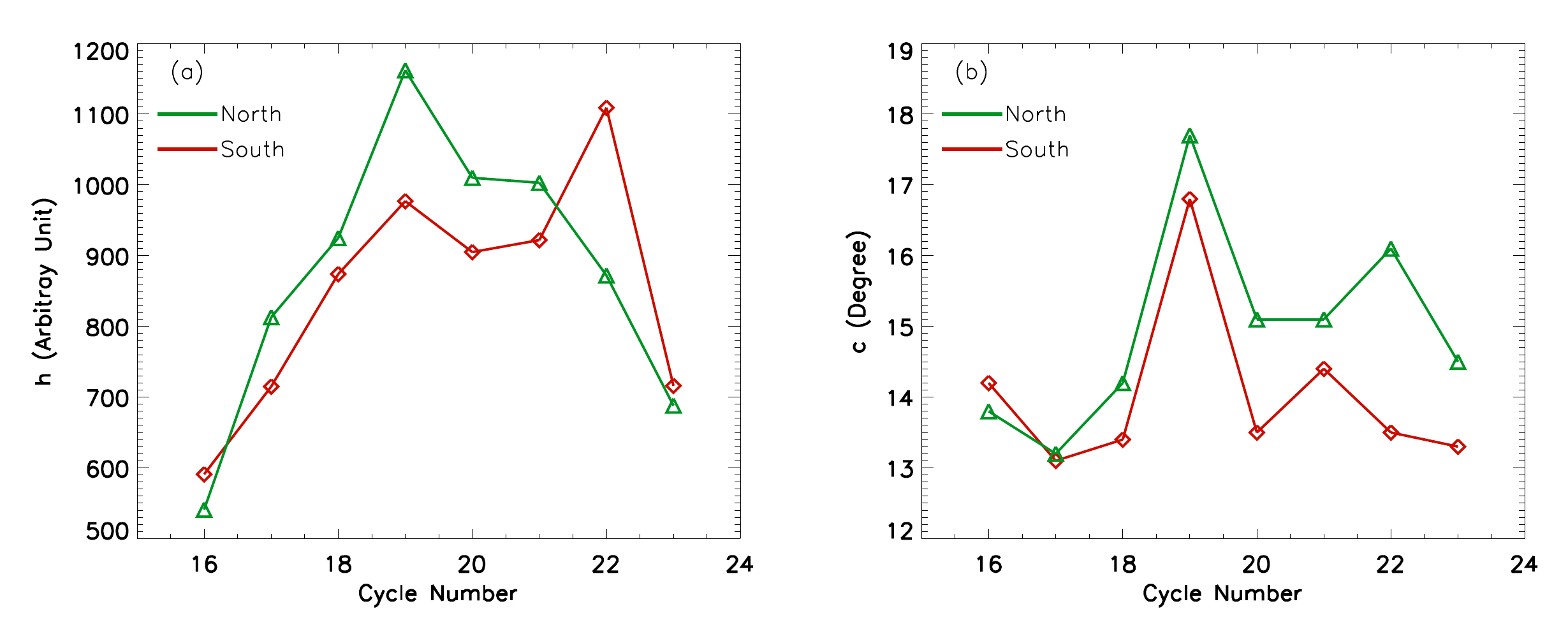}
\caption{ Panel (a) shows the height (h) of the fitted Gaussian functions (shown in Figure~\ref{lat_distri}) for northern and southern hemispheres in green and red colors respectively. Panel (b) shows the center values (c) of these Gaussian.}
\label{lat_deviation} 
\end{figure*}

  Next we explore the distribution of the number of the detected sunspots with their latitude. Panel (a) in Figure~\ref{lat_distri} shows the histogram for the period 1921 to 2011. We clearly see a `bi-model' distribution with two `bell-shaped' curves separated around the zero latitude. Such distribution is also seen for the individual cycles (panel (b-i)). We use a bin of 2$^{\circ}$, in latitude, to incorporate the fact that a sunspot may change its latitude slightly during its lifetime. We fit each of the peaks with a Gaussian function as shown in red thick line in all the panels in Figure~\ref{lat_distri}. Maximum Height (h) and center (c) of each of the fitted Gaussian is printed in individual panels. Apart from the Gaussian distributions, we also notice from the histograms that the height of the Gaussian peaks, in the two hemispheres, are different . Panel (a) of Figure~\ref{lat_deviation} shows the variation of the parameter `h', for both the hemispheres, with the cycle number. From the figure, we see that for 
most of the cycles the northern hemisphere dominates over the south. It must be emphasized here that this plot is effectively another representation of the N-S asymmetry. From panel (b) of Figure~\ref{lat_deviation} we see that maximum sunspots appear at a latitude of $\approx$ 14.5 degree on each side. We also notice a strong deviation of this center values from the average, for the cycle 19 which is the strongest cycle of the 20$^{th}$ century. Recently \citet{2016A&A...591A..46C} have explained the behavior of these fitted Gaussian parameters using a solar dynamo model and also predicted the value of the turbulent diffusivity acting on the toroidal magnetic field.

\section{Reproducing irregular features of solar cycles}\label{sec_wal}

Amplitude and length of a solar cycle varies over cycle to cycle. Based on various century long sunspot observations , it has been found that these irregularities of the solar cycles do have some definite properties (e.g, Waldmeier effect and the correlations between decay rate and cycle amplitude). \citet{1935Wald} reported that during the rising phase of a solar cycle, the rise time is anti-correlated with the peak amplitude of that cycle i.e, the stronger cycles take less time to rise and vice-versa. There is also a strong correlation between the rising rate and the peak amplitude of a cycle. These relations are known as Waldmeier effect. \citet{KarakChou11} reproduced these correlations using Greenwich sunspot data and classified these two properties as Waldmeier Effect 1 (WE1) and Waldmeier Effect 2 (WE2) respectively. During the descending phase of a solar cycle, some important correlations are also reported. Recently, using the Greenwich sunspot data, \citet{HKBC15} found a correlation between decay 
rate and the peak amplitude of the cycle. These authors have also shown that, apart from the sunspot area data, such correlation also exist in sunspot number and 10.7 cm radio flux data. Another important property during the decaying phase is the correlations between decay rate at late phase of a cycle and the amplitude of the succeeding cycle \citep{Yoshida10,HKBC15}. This is also important in terms of predicting the next cycle amplitude. Though this correlation is statistically not so strong, but it still gives an approximate estimation of the next cycle amplitude. \citet{HKBC15} have interpreted these relations in terms of a flux transport Dynamo model.

In this section we have calculated various correlations during ascending and descending phase of the solar cycles from the Kodaikanal sunspot area data and reproduced all the irregular features of the solar cycles discussed above.

\subsubsection{Waldmeier Effect: The correlations in ascending phase of the solar cycle}
\begin{figure*}[!htbp]
\centering
\includegraphics[width=.85\textwidth]{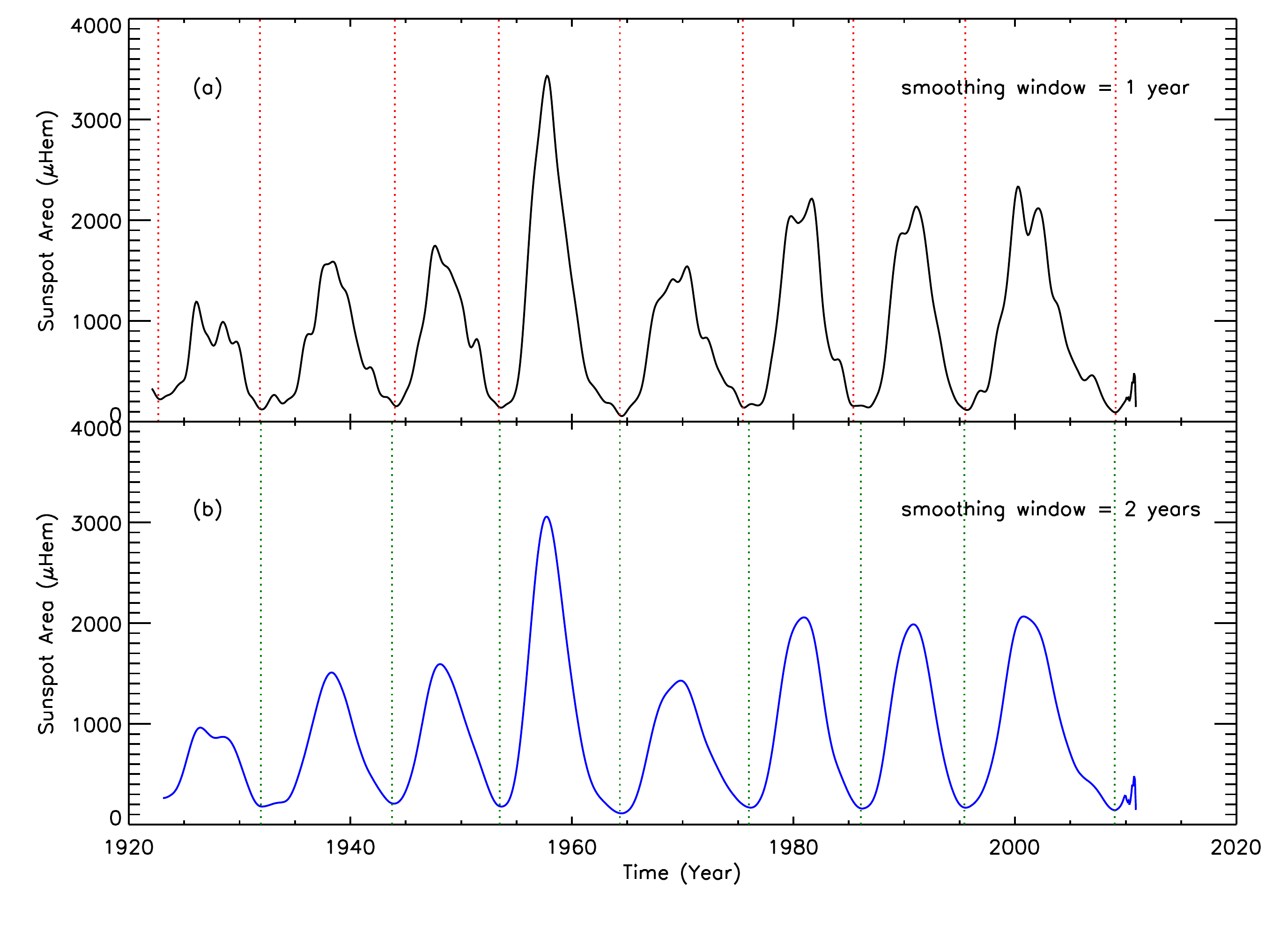} 

\caption{Panel (a) shows the smoothed (with a Gaussian of FWHM=1 year) solar cycle from kodaikanal sunspot area data. Panel (b): Same as panel (a) but with a FWHM of 2 years. }

\label{avg_1_2}
\end{figure*}

\begin{figure*}[!htbp]
\centering
\includegraphics[width=.85\textwidth]{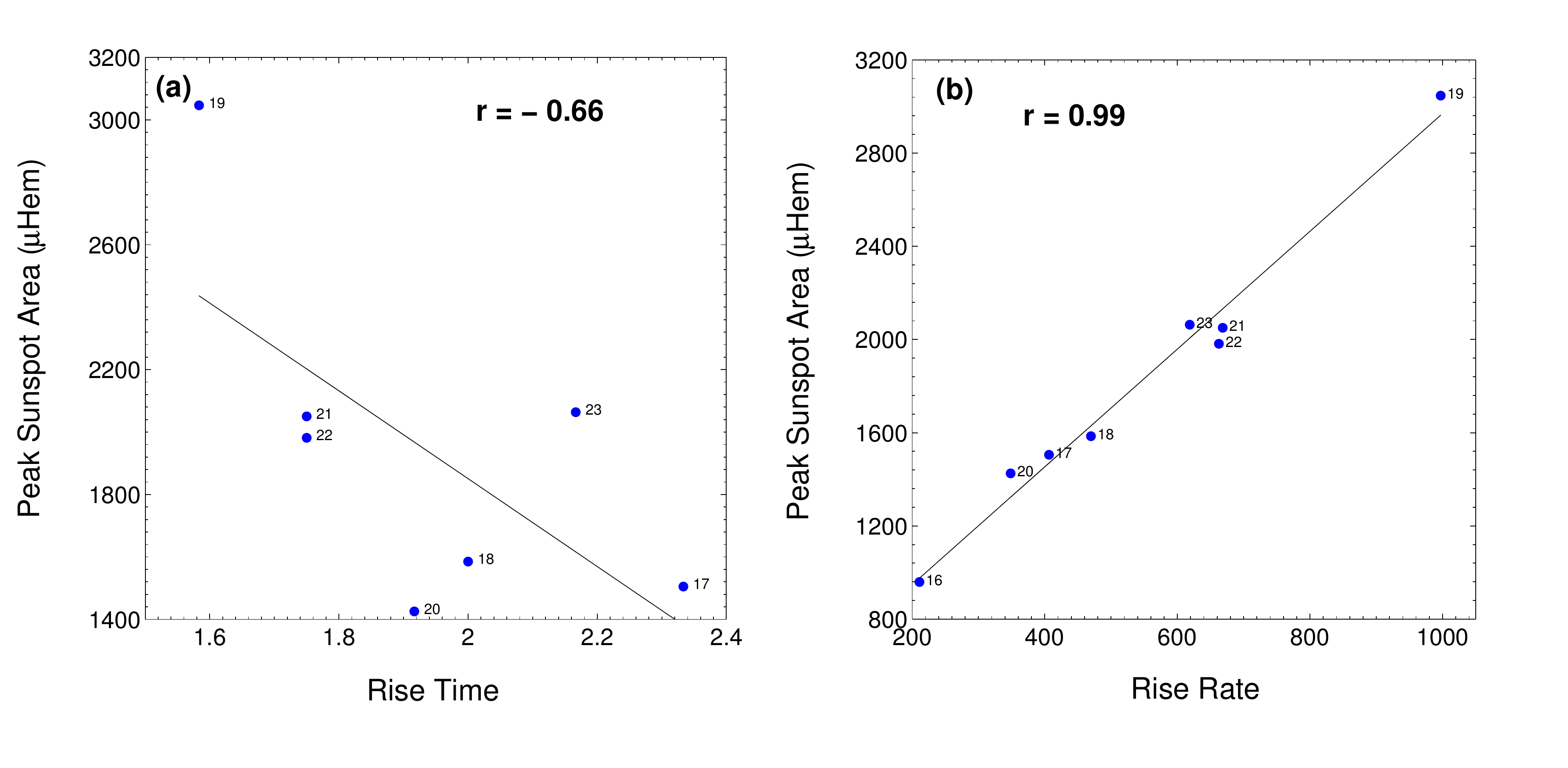} 

\caption{ Panel (a) shows the scatter plot of rise time (in years) and the peak amplitude of same solar cycle (FWHM=2 year). Solid line represents best linear fit. Scatter plot of rise rate ($\mu$Hem/year) and the peak amplitude of same solar cycle (FWHM=2 year) is shown in panel (b). Solar cycle numbers also marked beside each of the blue circles. }

\label{we11}
\end{figure*}

Although the irregular properties of the solar cycles are well established but doubts are always expressed about their existence. Are they really properties of
solar cycles or just statistical artifacts \citep{dikpati07}? Reconfirmation
of Waldmeier effect from a completely different sunspot area data set is not only
important for the prediction of the peak amplitude when cycle is in the early
stage but also it is important to validate our existing dynamo models. Time variation of daily sunspot area data or even the monthly averaged sunspot area data are so scattered that it is very difficult to calculate the required quantities like rise time and rise rate to validate their specific properties. Hence, we have smoothed our data by using a Gaussian filter having FWHM =1 year and FWHM = 2years as shown in Figure~\ref{avg_1_2}(a) and Figure~\ref{avg_1_2}(b) respectively. Red and green vertical dotted lines represent the  position of the minima calculated for one and two years averaged data sets. A closer inspection of Figure~\ref{avg_1_2} reveals that the determination of the solar minima for each cycle is always very sensitive to the filtering window and minima calculated using 1 year averaged data is not exactly same with the minima of the 2 years averaged data. This is true for determination of solar maxima also. It is also found that there is always an overlap between two successive cycles and the position of cycle minima depends on this overlapping \citep{CS07}.
 Thus, defining the solar cycle minima and the maxima becomes very difficult. Following \citet{KarakChou11}, we have calculated the rise time by taking the time difference between which the solar cycle reaches to $0.8$P from $0.2$P, where P = peak amplitude of the cycle. On the other hand rise rate is defined as the slope between two points separated by one year and the first point is chosen one year before the cycle maxima. All the plots shown in this section, has been generated using a dataset smoothed by a Gaussian filter with a FWHM of 2 years. Panel (a) in Figure~\ref{we11}  shows the plot  between the rise time and the peak amplitude of the cycle. We see an anti-correlation $(r = -0.66)$ between these two parameters and this effect is known as the WE1. Cycle 16, in panel (a) of Figure~\ref{we11}, has not been considered because the minima of this cycle is not well captured after two years of averaging. Therefore the position of the $0.2$P is not well defined in this case.
In panel (b) in Figure~\ref{we11}, the correlation between the rise rate and cycle amplitude is calculated and it is evident from the figure that WE2 is also well reproduced with correlation coefficients $r = 0.99$. Here we should mention that the correlation coefficient, which we have calculated here, is statistically significant only for large number of data points. In our case, we have only 8 data points corresponding to 8 cycles and one should calculate the ` adjusted Pearson correlation coefficient' ($r_{adj}$) defined as

\begin{equation}
\centering
r^2_{adj} = 1 - (1 - r^2)\frac{n-1}{n-p-1}
\end{equation}

where r is the normal correlation coefficient, n is the number of data points and p is the independent variables. The calculated adjusted coefficients for $r = -0.66$ (WE1) and $r = -0.99$ (WE2) become $r_{adj} = -0.57$ and $r_{adj} = 0.99$ respectively. Thus it shows that the `adjusted correlation coefficients' do not vary much for high value of correlation coefficients and thus we continue with the normal Pearson correlation coefficients.

In Table~\ref{gopalt} we have summarized the different correlation coefficient values which are calculated using Gaussian with a FWHM of 1 year and 2 years respectively.

\subsubsection{The correlations in descending phase of the solar cycle}

 In the previous section we mentioned that the presence of small fluctuations and the double peaks in the solar cycle makes it difficult to identify the actual minima and maxima of a cycle. To avoid this discrepancy, we have excluded the actual cycle minima and the maxima in our decay rate calculation. The final decay rate of a particular cycle has been determined by averaging over the individual decay rates calculated at four different locations. The first individual decay rate is determined by taking a slope between two points separated by one year and the first point is taken one year before the cycle minima. Similarly the last individual decay rate near solar maxima is also calculated by taking two points with one year interval and the first point is chosen one year after the solar maxima. The other two decay rates are calculated at locations which are in between positions of last and first decay rate points. 

\begin{figure*}[!htbp]
\centering
\includegraphics[width=.85\textwidth]{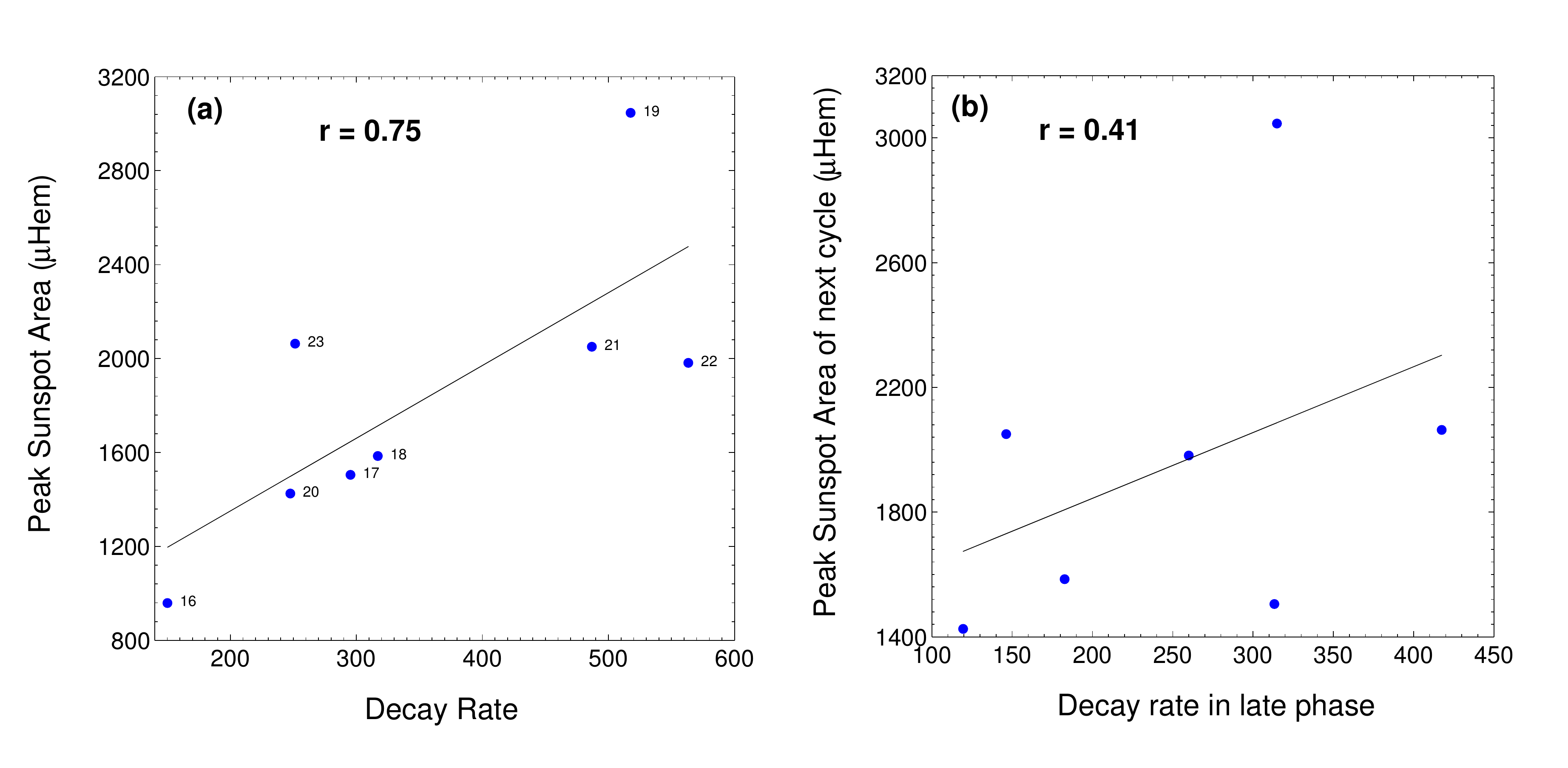}

\caption{ Panel (a) shows the scatter plot of decay rate and the peak amplitude of same solar cycle (FWHM=2 year). Best linear fit is shown with the solid line. Panel (b) shows the scatter plot of decay rate at late phase and the peak amplitude of the next cycle.} 

\label{decay1}
\end{figure*}
 In panel (a) of Figure~\ref{decay1} we have shown the correlations between the final decay rate of a cycle with the cycle amplitude using 2 years average sunspot area data. The correlations of decay rate calculated at the late phase of cycle with the amplitude of the next cycle is shown in panel (b) of Figure~\ref{decay1}.  We obtain a correlation coefficients $r= 0.75$ for the correlation between decay rate and cycle amplitude for same cycle whereas we get $r = 0.41$ for decay rate at late phase with next cycle amplitude. There are only seven points (cycle 23 is omitted) in panel (b) of Figure~\ref{decay1} due to the fact that the correlation is defined with the amplitude of the next cycle with the decay rate at the late phase of the previous cycle. The correlation coefficients obtained from one year averaged data (FWHM = 1 year) is also given in Table 1. From the values given in the table 1, we notice that only the correlation between rise time and the peak amplitude of the cycles is sensitive to the smoothing window (though statistical trend is same) whereas the other coefficients are of comparable values.

  We mentioned that each cycle is different from the previous in terms of cycle amplitude and duration. In the next section, we explore the periodic or quasi-periodic nature of these variations using the wavelet analysis.

\begin{table*}[!htbp]
\centering
\caption{Correlation coefficients at different phase of solar cycle}
\begin{tabular}{c|c|c|c|c}
\hline
Smoothed & \multicolumn{2}{|c|}{Correlations during rising phase}&\multicolumn{2}{|c|}{Correlations during decaying phase}\\
data with & \multicolumn{2}{|c|}{of the cycle between}&\multicolumn{2}{|c|}{of the cycle between}\\
\cline{2-5}
FWHM&Rise time and cycle & Rise rate and cycle & Decay rate and & Decay rate at late \\
&amplitude (WE1) & amplitude (WE2) &cycle amplitude &  phase and cycle amplitude\\
& & & & of next cycle\\
\hline
1 year & -0.63 & 0.92 & 0.57 & 0.56   \\
&&&&\\
2 years & -0.66 & 0.99 & 0.75 & 0.41 \\ 
\hline
\end{tabular}
\label{gopalt}
\end{table*} 
 

\section{Periodicities in the sunspot area data}\label{sec_period}

Sunspot cycle has an average periodicity of 11-yeas, though there are also various reports of periodicities shorter than 11 years~\citep{lrsp-2015-4}. Two of the most significant and persistent periods (shorter than 11-year period), seem to be around 155 days and 1.3 years \citep{1990A&A...238..377C,2002A&A...394..701K}. Though the origin of the 11-year sunspot cycle is now believed to be governed by the global solar dynamo mechanism but the origin of these shorter periodicities are not very clearly understood till now. 
 
\begin{figure*}[!htbp]
\centering
\includegraphics[angle=90,trim = 0mm 0mm 51mm 0mm, clip,width=.90\textwidth]{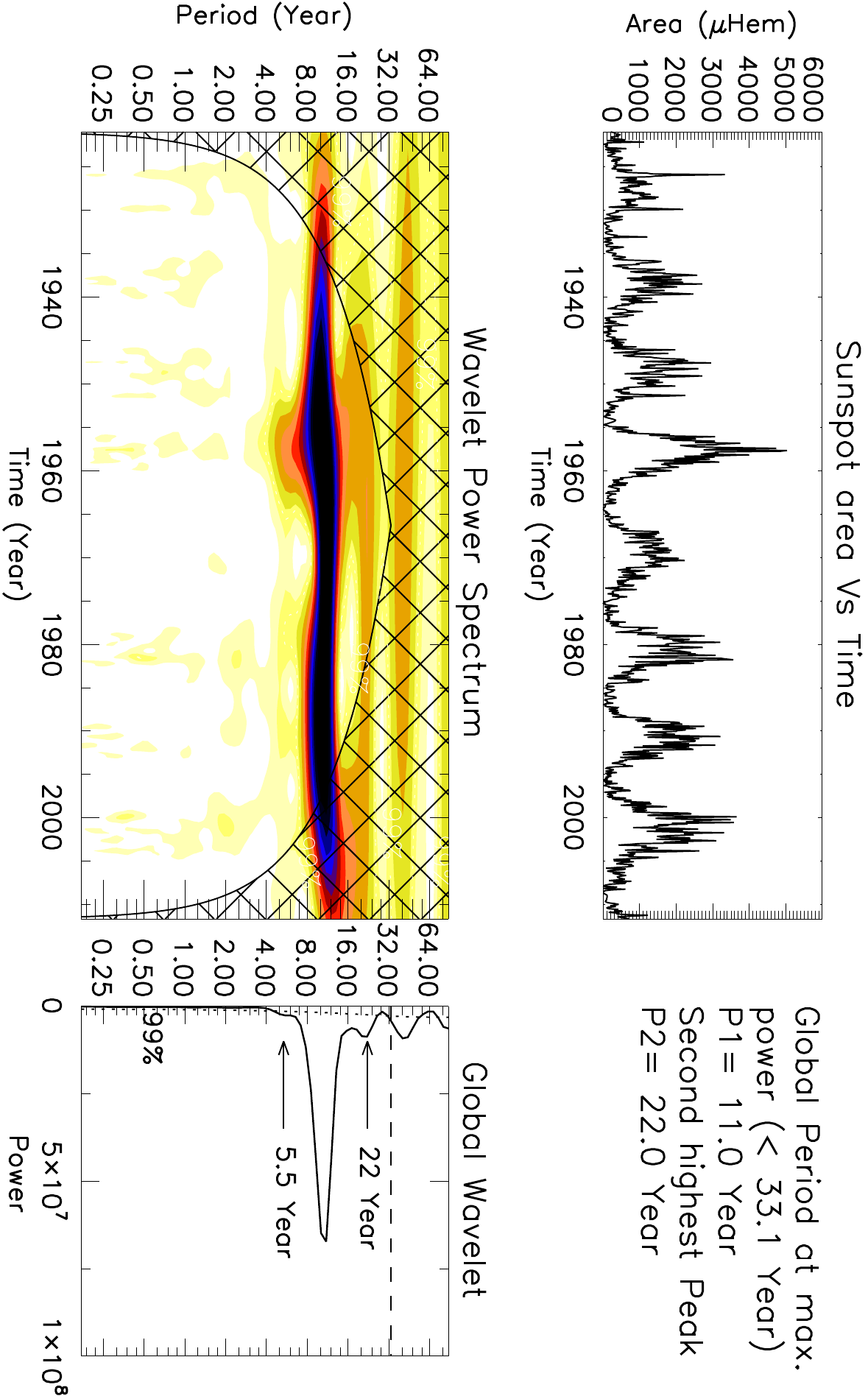}
\caption{The top panel shows the variations of monthly averaged sunspot area data, from Kodaikanal, with time. 
The bottom left panel shows the wavelet power spectrum. Cross-hatched regions above the wavelet power spectrum is the cone of influence (COI).
The bottom right panel shows the global wavelet power.
The maximum measurable period is $\sim$33 year which is shown by a horizontal dashed line. The dotted line
above the global wavelet power plot shows the significance level of 99~\%. The significant periods as measured from the global wavelet power is printed on the panel. 
}
\label{fig:11y}
\end{figure*}

In this section, we study the periodic variations of different time scales using the monthly averaged sunspot area data obtained from the Kodaikanal Observatory.
We use the wavelet analysis tool \citep{1998BAMS...79...61T} to study the periodic and quasi-periodic variations in the sunspot area time series. 
The wavelet analysis is an useful tool to examine the presence of localized oscillations, both in time and frequency domains. Figure~\ref{fig:11y} shows 
the wavelet analysis on the monthly averaged sunspot area time series. The top panel shows the time variations of the monthly averaged sunspot area whereas the bottom 
left panel shows the wavelet power spectrum. The global wavelet power, defined as the time-averaged wavelet power, is shown in the bottom right panel. 
The dotted line on top of the global wavelet power shows the significance level of 99~\% which is calculated by assuming a white noise \citep{1998BAMS...79...61T}. 
The global wavelet power shows a dominant period of 11-years with the presence of weak power concentrations at 22 and 5.5 years periods. 
The 11-year period is the strongest period present in the sunspot activity and probably governed by the global
dynamo mechanism (see \cite{lrsp-2010-3}). 

Short term periodicities, between 1 year to $\approx$ 4 years, are often referred as quasi-biennial oscillations (QBOs). QBOs were previously found in different solar proxies such as, in solar wind data, in cosmic rays etc (see Table~1 in \citet{2014SSRv..186..359B} for a complete list). QBOs are also observed in the frequency shifts as inferred from the helioseismology data ( \citet{2015SoPh..290.3095B}). Periodicities in such variety of proxies suggest that these small period oscillations are a real effect and have an origin connected to the solar dynamo.

\begin{figure*}[!htbp]
\centering
\includegraphics[angle=90,trim = 0mm 0mm 51mm 0mm, clip,width=.90\textwidth]{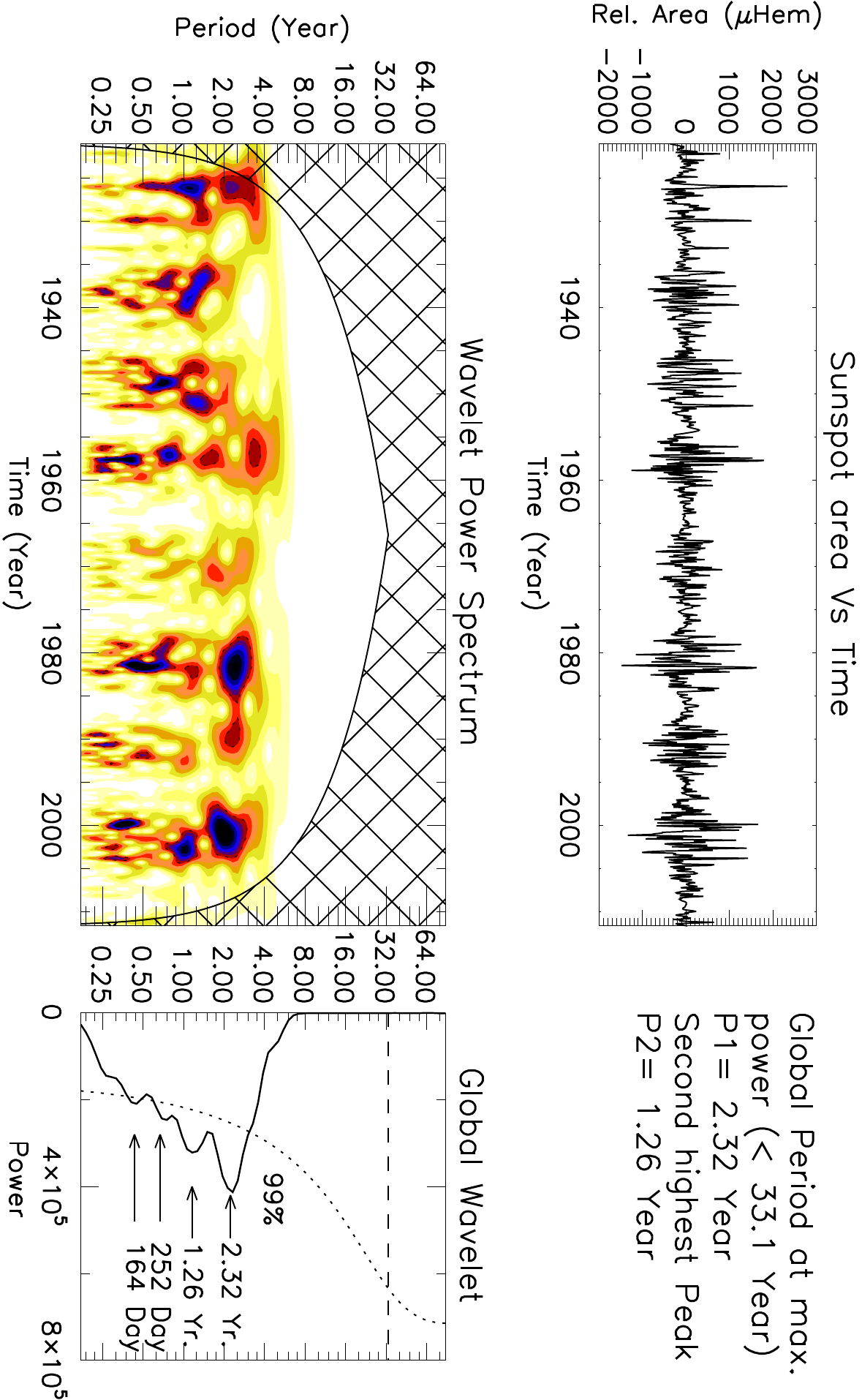}
\caption{The top panel shows the variations of sunspot area with time after removing low-frequency components form the time series.
The different panels are same as in Figure~\ref{fig:11y}.} 
\label{fig:1p3y}
\end{figure*}
\begin{figure*}[!htbp]
\centering
\includegraphics[angle=90,trim = 0mm 0mm 51mm 0mm, clip,width=.90\textwidth]{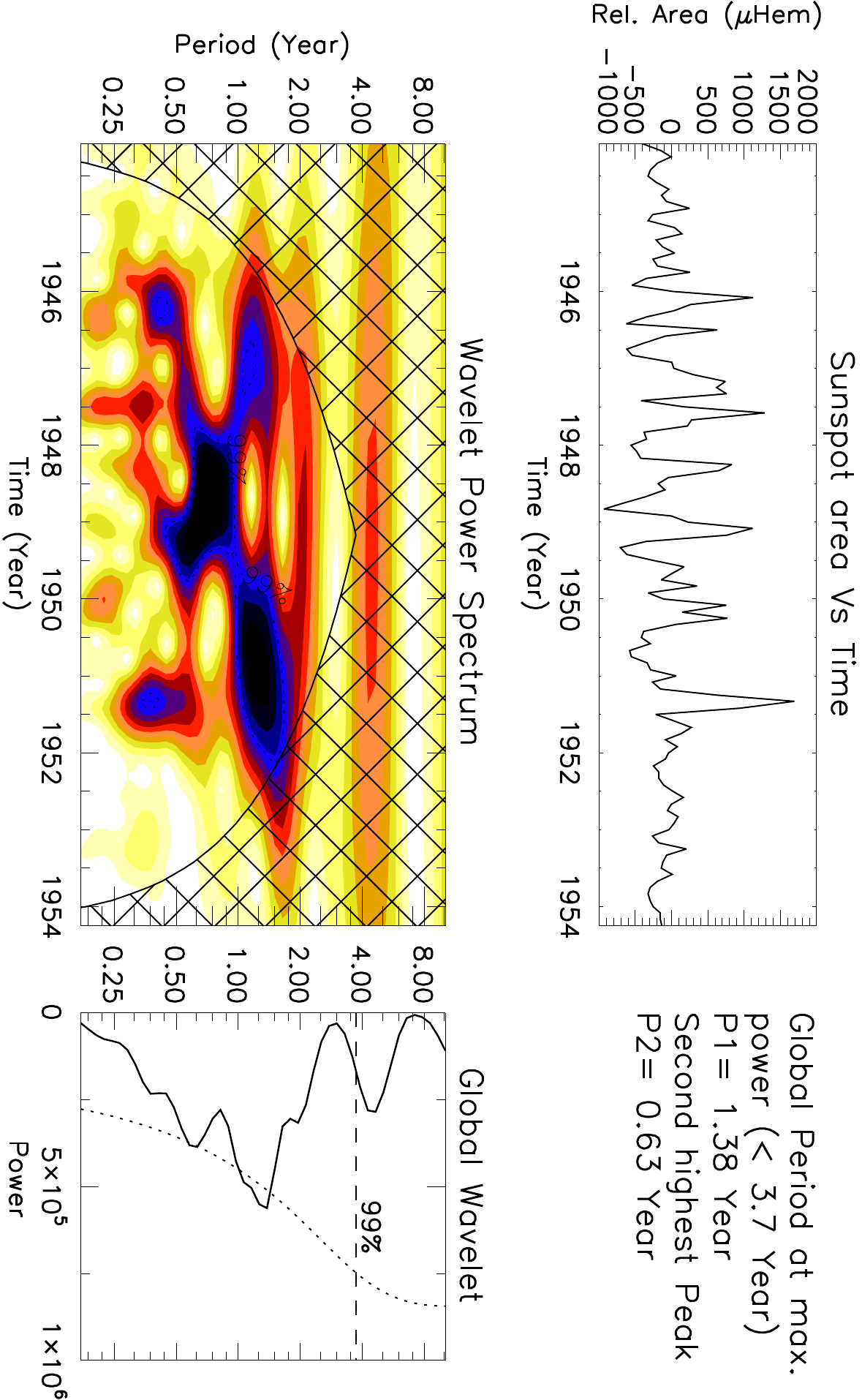}
\caption{The top panel shows the variations of sunspot area with time (solar cycle 18) after removing the low-frequency components form the time series.
The panels are as in Figure~\ref{fig:11y}} 
\label{fig:cycle18}
\end{figure*}

The 22-year cycle is probably a manifestation of the Gnevyshev-Ohl (even-odd) rule, which refers to a pattern of alternating higher and lower cycle amplitudes than the average cycle amplitude \citep{1948..Astron..Zh..25..18G,2007ApJ...658..657C}.  The 5.5-year period, apart from being a harmonic of the 11-year cycle, is closely related to the double peak behavior of the solar cycle \citep{1967SoPh....1..107G,1977SoPh...51..175G,2011ISRAA2011E...2G}.

The sunspot cycle is highly dominated by the 11-year periodicity. To explore the presence of the short-term periodicities in the time series, one needs to suppress the 11-year period from the time series. To remove the longer periods, we used the empirical mode decomposition (EMD) technique \citep{1998RSPSA.454..903E}. 
 Time series, after removing the longer periods, is shown in the top panel of the Figure~\ref{fig:1p3y}. We have filtered out the periods longer than $\sim$4 years to find the periodic and/or quasi-periodic variations within the solar cycles. The results of the wavelet analysis is shown in Figure~\ref{fig:1p3y}. It shows that the shorter-period oscillations are evolving very rapidly compared to the long-period oscillations. It also shows that the shorter-period oscillations are prominent around the solar maximum (similar to \citet{2006ChA&A..30..393Z}). The global wavelet plot shows significant powers at 2.32-year, 1.26-year, 252-day and 164-day. The presence of 150--160 day periodicity in the solar activity is widely reported \citep{1989ApJ...337..568L,1990ApJ...363..718L,1990A&A...238..377C,1998Natur.394..552O,2002A&A...394..701K,2002ApJ...566..505B}
along with the 1.3-year period \citep{2001JGR...10616021L,2002A&A...394..701K,2007AdSpR..40.1006O}. \citet{2006ChA&A..30..393Z} found the existence of 2.21-year and 246-day periodicity in the sunspot area data obtained from the Royal Greenwich Observatory. We have also analyzed the individual cycles separately and the obtained periodicities are listed in table~\ref{tab10}. One such case, for cycle 18, is shown in Figure~\ref{fig:cycle18}.

\begin{table*}[!htbp]
\centering
\caption{Periodicities in the individual solar cycles}
\begin{tabular}{c|c|c|c}

\hline
\hline
\multicolumn{1}{c}{Cycle Number}&\multicolumn{1}{c}{Strongest Period}&\multicolumn{1}{c}{Second strongest Period}&\multicolumn{1}{c}{Third strongest period}\\
\hline
\hline
 16 & 3.26~Y & 1.26~D & 230~D  \\
 17 & 1.26~Y & 2.32~Y & 164~D  \\
 18 & 1.38~Y & 230~D  & 1.95~Y \\
 19 & 1.95~Y & 274~D & 3.28~Y  \\
 20 & 3.28~Y & 1.64~Y & 164~D  \\
 21 & 2.76~Y & 252~D & 164~D  \\
 22 & 3.00~Y & 1.95~Y & 1.06~Y \\
 23 & 1.95~Y & 354~D & 124~D \\
\hline
\hline
\end{tabular}
\label{tab10}
\end{table*}


\section{Summary and conclusion}\label{sec_summary}

In this paper, we presented a century long white-light digitized data from Kodaikanal observatory. The highlight of this data set is that the data have been acquired with the same optics since 1918. Such a consistent data set is extremely useful for long term variations in the solar activities. We have implemented a semi-automated algorithm for the sunspot detection throughout this dataset. Thus, it minimizes the effect of human bias on the detection method and enhances the confidence on the results obtained from it. 

   We have presented the sunspot area variation over different solar cycles and found a very good correlation with Greenwich data. Sunspot area values, obtained using hand drawn contours on the Kodaikanal white-light images, by \citet{2000JApA...21..149S} match well with our  semi-automated detection. 
   
   We found that the sunspot sizes follow an exponential distribution for each solar cycle. Such distribution do not change at different phases of the solar cycle though the number of sunspots with different area values seem to increase or decrease according to the cycle maxima or cycle minima.

We have measured the hemispheric asymmetry associated with the sunspot area. Comparing the two popular methods of ascribing the asymmetricity, we have shown that the two methods give different results in-term of peak time of these parameters. Also the double peak behavior seen the solar cycles often comes from the double peaks seen in one hemisphere only. Latitudinal distribution of the detected sunspot shows that most of the sunspots appear at $\sim$ 15$^{\circ}$ latitude in each of the hemispheres with a Gaussian like distribution function \citep{2011A&A...528A..82J}. The peak heights of the Gaussian distributions, for each hemisphere for a particular cycle, are different and shows a long term variation. From the `butterfly diagram', we show that the bigger sunspots (area$\geq$ 1000$\mu$Hem) appear  closer to solar maxima and their numbers are directly proportional to the cycle strength, whereas no such relationship exists for the smaller sunspots (area$\leq$ 30$\mu$Hem). Using, this area 
time series we have confirmed the `Waldmeier effect' which relates the rise or decay rate of cycle with the current and next cycle amplitudes. To find the different periodicities present in the individual solar cycles we have performed wavelet analysis. We have found presence of  periodicities as long as 22 years to as small as 124 days.

The entire time series (1921 -- 2011) from Kodaikanal including the  area, latitude and longitude information corresponding to each of the detected sunspots is provided as a supplementary material (available online). A separate list containing the spotless days is also available online. We are planning to complete the calibration process of the initial 15 years (from 1905 to 1920) in future. The time series will then be updated ( with proper version number) as soon as we complete the process.

\begin{acknowledgements}

We would like to thank all the observers at Kodaikanal over 100 years for their contribution to build this enormous resource. The current high resolution digitization process was initiated by Prof. Jagdev singh and we thank him for his enormous contribution in the project. We would like to thank many who has helped in the digitization and calibration : Muthu Priyal, Amareswari, T. G. Priya, Ayesha Banu, A. Nazia, S. Kamesh, P. Manikantan, Janani, Trupti Patil and Sudha. Staff members at Kodaikanal also helped us in setting up the digitizer unit and the digitization at Kodaikanal. The authors would also like to thank Prof Arnab Rai Choudhuri for his valuable comments which helped us to improved the quality of the paper. Finally we would like to thank the anonymous referee for his/her valuable comments and suggestions, which improved the quality of the presentation.
\end{acknowledgements}

\textbf{Appendix A}\label{app}

Sunspot area distribution had been studied in the past by \citet{1988ApJ...327..451B,2005A&A...443.1061B}. These authors have used a lognormal distribution function to fit the area data. We also follow the same and fitted the data with a lognormal function defined as

\hspace{5cm}$\mathrm{y}=\frac{1}{\sqrt{2\pi}\sigma x}\exp{-\frac{[\ln(x)-\mu]^2}{2\sigma^2.}}$

\begin{figure*}[!htbp]
\centering
\includegraphics[angle=90,width=1.\textwidth]{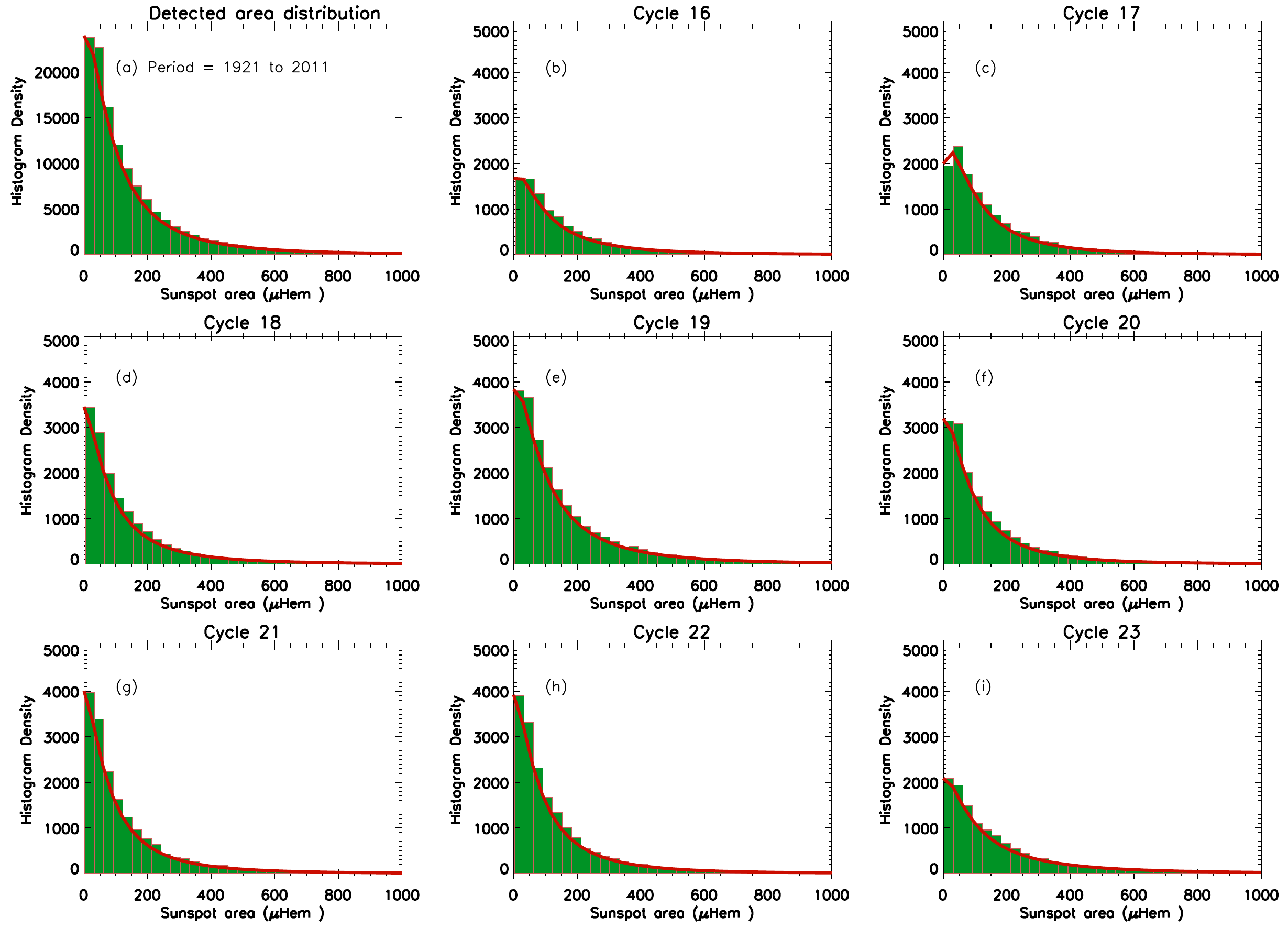}
\caption{ Plot showing the histograms (with a bin-size of 30 $\mu$Hem) of the individual area values (sizes) obtained from Kodaikanal white-light images. The histogram for the full data set is shown in panel (a). Panels (b-i) show the area distributions for the individual solar cycles ( cycle 16 to 23). }
\label{log_plot} 
\end{figure*}

 Where $\mu$ and $\sigma$ are the mean and the standard deviation respectively. The fit on the histograms of the sunspot area distributions, for the full period and for the individual cycles, are shown in Figure~\ref{log_plot}. Here we must emphasize the fact that the lognormal distribution first increases and then decreases exponentially as $\exp(\mu-\sigma^2)$. Since the initial increment of the lognormal distribution is restricted very close to the origin of the histogram, so it almost mimics the exponentially decaying distribution as shown previously in Figure\ref{area_hist}.

\begin{figure*}[!htbp]
\centering
\includegraphics[width=0.5\textwidth]{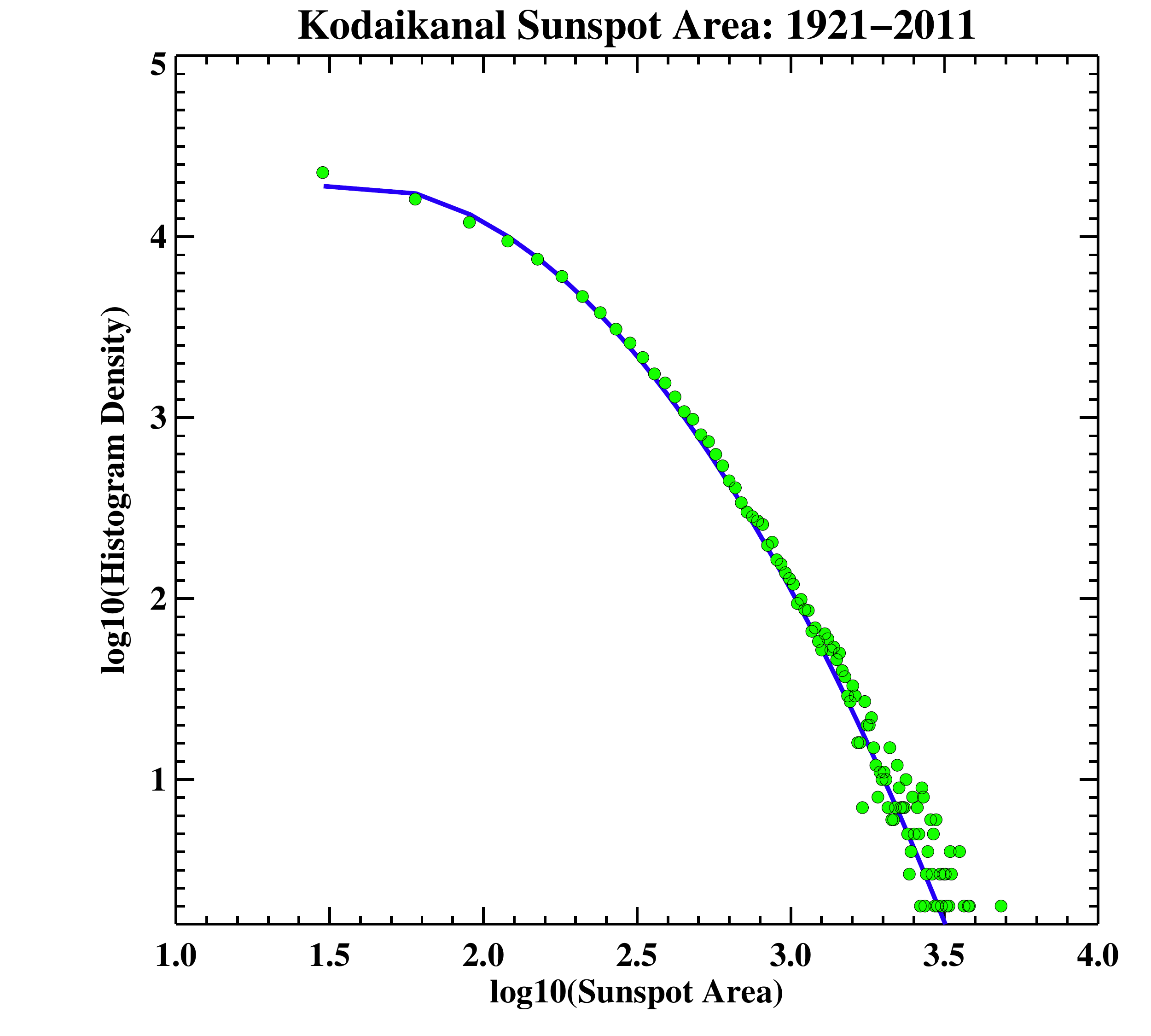}
\caption{ Same as Panel (a) of Figure~\ref{log_plot} but in a log-log scale. The blue solid line shows the fitted lognormal function.}
\label{log_log_plot} 
\end{figure*}

  Also, we plot the same quantity as in Panel (a) of Figure~\ref{log_plot} but in a natural logarithmic scale in Figure~\ref{log_log_plot}. We also fit data with a lognormal function as shown with the solid blue line in the figure. The lognormal function appears as a parabola in a log-log space.

\textbf{Appendix B}

 We also compare our Kodaikanal sunspot area values with the Debrecen data \citep{2011IAUS..273..403G,2016SoPh..tmp..124B} which covers the period from 1974 to 2016 (and continuing). Though some of the gaps in the data has been filled up by using data from other observatories, but the catalog is mainly based on Debrecen and Gyula full-disk white-light observations.
\begin{figure*}[!htbp]
\centering
\includegraphics[angle=90,clip,width=.95\textwidth]{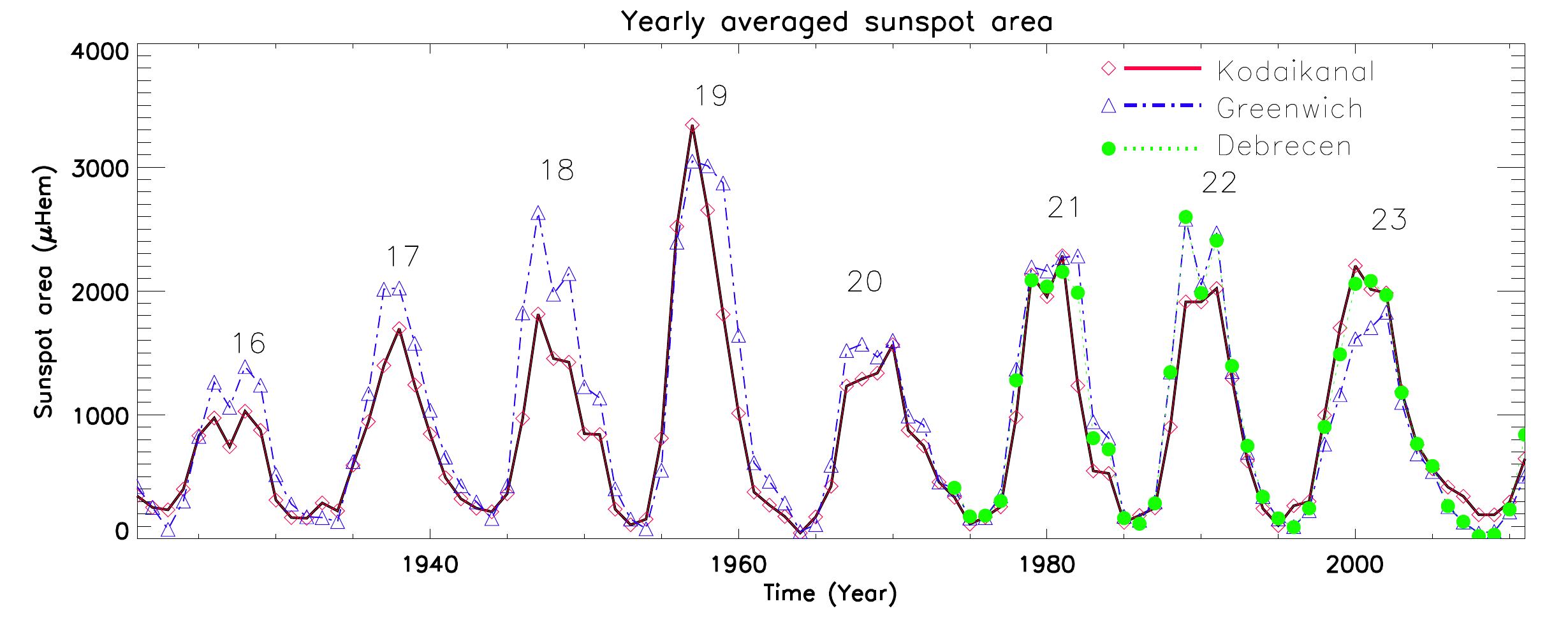}
\caption{ Yearly averaged sunspot area values from Kodaikanal and Greenwich and Debrecen sunspot catalogs.}
\label{deb_image} 
\end{figure*}
We further compute the correlation coefficient between the Kodaikanal and the Debrecen data sets and it is equal to 0.97.

\bibliographystyle{aa}

\begin{thebibliography}{49}
\expandafter\ifx\csname natexlab\endcsname\relax\def\natexlab#1{#1}\fi

\bibitem[{{Badalyan} \& {Obridko}(2011)}]{2011NewA...16..357B}
{Badalyan}, O.~G. \& {Obridko}, V.~N. 2011, \na, 16, 357

\bibitem[{{Ballester} {et~al.}(2002){Ballester}, {Oliver}, \&
  {Carbonell}}]{2002ApJ...566..505B}
{Ballester}, J.~L., {Oliver}, R., \& {Carbonell}, M. 2002, \apj, 566, 505

\bibitem[{{Ballester} {et~al.}(2005){Ballester}, {Oliver}, \&
  {Carbonell}}]{2005A&A...431L...5B}
{Ballester}, J.~L., {Oliver}, R., \& {Carbonell}, M. 2005, \aap, 431, L5

\bibitem[{{Bappu}(1967)}]{1967SoPh....1..151B}
{Bappu}, M.~K.~V. 1967, \solphys, 1, 151

\bibitem[{{Baranyi} {et~al.}(2016){Baranyi}, {Gy{\H o}ri}, \&
  {Ludm{\'a}ny}}]{2016SoPh..tmp..124B}
{Baranyi}, T., {Gy{\H o}ri}, L., \& {Ludm{\'a}ny}, A. 2016, \solphys
  [\eprint[arXiv]{1606.00669}]

\bibitem[{{Baumann} \& {Solanki}(2005)}]{2005A&A...443.1061B}
{Baumann}, I. \& {Solanki}, S.~K. 2005, \aap, 443, 1061

\bibitem[{{Bazilevskaya} {et~al.}(2014){Bazilevskaya}, {Broomhall}, {Elsworth},
  \& {Nakariakov}}]{2014SSRv..186..359B}
{Bazilevskaya}, G., {Broomhall}, A.-M., {Elsworth}, Y., \& {Nakariakov}, V.~M.
  2014, \ssr, 186, 359

\bibitem[{{Bogdan} {et~al.}(1988){Bogdan}, {Gilman}, {Lerche}, \&
  {Howard}}]{1988ApJ...327..451B}
{Bogdan}, T.~J., {Gilman}, P.~A., {Lerche}, I., \& {Howard}, R. 1988, \apj,
  327, 451

\bibitem[{{Broomhall} \& {Nakariakov}(2015)}]{2015SoPh..290.3095B}
{Broomhall}, A.-M. \& {Nakariakov}, V.~M. 2015, \solphys, 290, 3095

\bibitem[{{Cameron} \& {Sch{\"u}ssler}(2007)}]{CS07}
{Cameron}, R. \& {Sch{\"u}ssler}, M. 2007, \apj, 659, 801

\bibitem[{{Cameron} \& {Sch{\"u}ssler}(2016)}]{2016A&A...591A..46C}
{Cameron}, R.~H. \& {Sch{\"u}ssler}, M. 2016, \aap, 591, A46

\bibitem[{{Carbonell} \& {Ballester}(1990)}]{1990A&A...238..377C}
{Carbonell}, M. \& {Ballester}, J.~L. 1990, \aap, 238, 377

\bibitem[{{Carbonell} {et~al.}(1993){Carbonell}, {Oliver}, \&
  {Ballester}}]{1993A&A...274..497C}
{Carbonell}, M., {Oliver}, R., \& {Ballester}, J.~L. 1993, \aap, 274, 497

\bibitem[{Charbonneau(2010)}]{lrsp-2010-3}
Charbonneau, P. 2010, Living Reviews in Solar Physics, 7

\bibitem[{{Charbonneau} {et~al.}(2007){Charbonneau}, {Beaubien}, \&
  {St-Jean}}]{2007ApJ...658..657C}
{Charbonneau}, P., {Beaubien}, G., \& {St-Jean}, C. 2007, \apj, 658, 657

\bibitem[{{Dikpati} {et~al.}(2008){Dikpati}, {Gilman}, \& {de
  Toma}}]{dikpati07}
{Dikpati}, M., {Gilman}, P.~A., \& {de Toma}, G. 2008, \apjl, 673, L99

\bibitem[{{Georgieva}(2011)}]{2011ISRAA2011E...2G}
{Georgieva}, K. 2011, ISRN Astronomy and Astrophysics, 2011, 437838

\bibitem[{{Gnevyshev}(1967)}]{1967SoPh....1..107G}
{Gnevyshev}, M.~N. 1967, \solphys, 1, 107

\bibitem[{{Gnevyshev}(1977)}]{1977SoPh...51..175G}
{Gnevyshev}, M.~N. 1977, \solphys, 51, 175

\bibitem[{{Gnevyshev} \& {Ohl}(1948)}]{1948..Astron..Zh..25..18G}
{Gnevyshev}, M.~N. \& {Ohl}, A.~I. 1948, Astron. Zh., 25, 18

\bibitem[{{Gupta} {et~al.}(1999){Gupta}, {Sivaraman}, \&
  {Howard}}]{1999SoPh..188..225G}
{Gupta}, S.~S., {Sivaraman}, K.~R., \& {Howard}, R.~F. 1999, \solphys, 188, 225

\bibitem[{{Gy{\H o}ri} {et~al.}(2011){Gy{\H o}ri}, {Baranyi}, \&
  {Ludm{\'a}ny}}]{2011IAUS..273..403G}
{Gy{\H o}ri}, L., {Baranyi}, T., \& {Ludm{\'a}ny}, A. 2011, in IAU Symposium,
  Vol. 273, Physics of Sun and Star Spots, ed. D.~{Prasad Choudhary} \& K.~G.
  {Strassmeier}, 403--407

\bibitem[{{Hasan} {et~al.}(2010){Hasan}, {Mallik}, {Bagare}, \&
  {Rajaguru}}]{2010ASSP...19...12H}
{Hasan}, S.~S., {Mallik}, D.~C.~V., {Bagare}, S.~P., \& {Rajaguru}, S.~P. 2010,
  Astrophysics and Space Science Proceedings, 19, 12

\bibitem[{Hathaway(2015)}]{lrsp-2015-4}
Hathaway, D.~H. 2015, Living Reviews in Solar Physics, 12

\bibitem[{{Hazra} {et~al.}(2015){Hazra}, {Karak}, {Banerjee}, \&
  {Choudhuri}}]{HKBC15}
{Hazra}, G., {Karak}, B.~B., {Banerjee}, D., \& {Choudhuri}, A.~R. 2015,
  \solphys, 290, 1851

\bibitem[{{Howard} {et~al.}(1999){Howard}, {Gupta}, \&
  {Sivaraman}}]{1999SoPh..186...25H}
{Howard}, R.~F., {Gupta}, S.~S., \& {Sivaraman}, K.~R. 1999, \solphys, 186, 25

\bibitem[{{Howard} {et~al.}(2000){Howard}, {Sivaraman}, \&
  {Gupta}}]{2000SoPh..196..333H}
{Howard}, R.~F., {Sivaraman}, K.~R., \& {Gupta}, S.~S. 2000, \solphys, 196, 333

\bibitem[{{Huang} {et~al.}(1998){Huang}, {Shen}, {Long}, {Wu}, {Shih}, {Zheng},
  {Yen}, {Tung}, \& {Liu}}]{1998RSPSA.454..903E}
{Huang}, N.~E., {Shen}, Z., {Long}, S.~R., {et~al.} 1998, Proceedings of the
  Royal Society of London Series A, 454, 903

\bibitem[{{Jiang} {et~al.}(2011){Jiang}, {Cameron}, {Schmitt}, \&
  {Sch{\"u}ssler}}]{2011A&A...528A..82J}
{Jiang}, J., {Cameron}, R.~H., {Schmitt}, D., \& {Sch{\"u}ssler}, M. 2011,
  \aap, 528, A82

\bibitem[{{Karak} \& {Choudhuri}(2011)}]{KarakChou11}
{Karak}, B.~B. \& {Choudhuri}, A.~R. 2011, \mnras, 410, 1503

\bibitem[{{Krivova} \& {Solanki}(2002)}]{2002A&A...394..701K}
{Krivova}, N.~A. \& {Solanki}, S.~K. 2002, \aap, 394, 701

\bibitem[{{Lean}(1990)}]{1990ApJ...363..718L}
{Lean}, J. 1990, \apj, 363, 718

\bibitem[{{Lean} \& {Brueckner}(1989)}]{1989ApJ...337..568L}
{Lean}, J.~L. \& {Brueckner}, G.~E. 1989, \apj, 337, 568

\bibitem[{{Lockwood}(2001)}]{2001JGR...10616021L}
{Lockwood}, M. 2001, \jgr, 106, 16021

\bibitem[{{Obridko} \& {Shelting}(2007)}]{2007AdSpR..40.1006O}
{Obridko}, V.~N. \& {Shelting}, B.~D. 2007, Advances in Space Research, 40,
  1006

\bibitem[{{Oliver} \& {Ballester}(1994)}]{1994SoPh..152..481O}
{Oliver}, R. \& {Ballester}, J.~L. 1994, \solphys, 152, 481

\bibitem[{{Oliver} {et~al.}(1998){Oliver}, {Ballester}, \&
  {Baudin}}]{1998Natur.394..552O}
{Oliver}, R., {Ballester}, J.~L., \& {Baudin}, F. 1998, \nat, 394, 552

\bibitem[{{Ravindra} {et~al.}(2013){Ravindra}, {Priya}, {Amareswari}, {Priyal},
  {Nazia}, \& {Banerjee}}]{2013A&A...550A..19R}
{Ravindra}, B., {Priya}, T.~G., {Amareswari}, K., {et~al.} 2013, \aap, 550, A19

\bibitem[{{Sivaraman}(2000)}]{2000JApA...21..149S}
{Sivaraman}, K.~R. 2000, Journal of Astrophysics and Astronomy, 21, 149

\bibitem[{{Sivaraman} {et~al.}(1993){Sivaraman}, {Gupta}, \&
  {Howard}}]{1993SoPh..146...27S}
{Sivaraman}, K.~R., {Gupta}, S.~S., \& {Howard}, R.~F. 1993, \solphys, 146, 27

\bibitem[{{Sivaraman} {et~al.}(1999){Sivaraman}, {Gupta}, \&
  {Howard}}]{1999SoPh..189...69S}
{Sivaraman}, K.~R., {Gupta}, S.~S., \& {Howard}, R.~F. 1999, \solphys, 189, 69

\bibitem[{{Sivaraman} {et~al.}(2003){Sivaraman}, {Sivaraman}, {Gupta}, \&
  {Howard}}]{2003SoPh..214...65S}
{Sivaraman}, K.~R., {Sivaraman}, H., {Gupta}, S.~S., \& {Howard}, R.~F. 2003,
  \solphys, 214, 65

\bibitem[{{Temmer} {et~al.}(2006){Temmer}, {Ryb{\'a}k}, {Bend{\'{\i}}k},
  {Veronig}, {Vogler}, {Otruba}, {P{\"o}tzi}, \&
  {Hanslmeier}}]{2006A&A...447..735T}
{Temmer}, M., {Ryb{\'a}k}, J., {Bend{\'{\i}}k}, P., {et~al.} 2006, \aap, 447,
  735

\bibitem[{{Torrence} \& {Compo}(1998)}]{1998BAMS...79...61T}
{Torrence}, C. \& {Compo}, G.~P. 1998, Bulletin of the American Meteorological
  Society, 79, 61

\bibitem[{{Vaquero}(2007)}]{2007AdSpR..40..929V}
{Vaquero}, J.~M. 2007, Advances in Space Research, 40, 929

\bibitem[{{Waldmeier}(1935)}]{1935Wald}
{Waldmeier}, M. 1935, Mitt. Eidgen. Sternw. Zurich, 14, 105

\bibitem[{{Watson} {et~al.}(2009){Watson}, {Fletcher}, {Dalla}, \&
  {Marshall}}]{2009SoPh..260....5W}
{Watson}, F., {Fletcher}, L., {Dalla}, S., \& {Marshall}, S. 2009, \solphys,
  260, 5

\bibitem[{{Yoshida} \& {Yamagishi}(2010)}]{Yoshida10}
{Yoshida}, A. \& {Yamagishi}, H. 2010, Ann. Geophysicae, 28, 417

\bibitem[{{Zhan} {et~al.}(2006){Zhan}, {He}, {Ye}, \&
  {Zhao}}]{2006ChA&A..30..393Z}
{Zhan}, L.-s., {He}, J.-m., {Ye}, Y.-l., \& {Zhao}, H.-j. 2006, Chinese
  Astronomy and Astrophysics,, 30, 393

\end{thebibliography}

\end{document}